\documentclass[useAMS]{gGAF2e}
\usepackage{amsmath}
\usepackage{amssymb}
\usepackage{graphicx}
\usepackage[authoryear]{natbib}

\providecommand{\tabularnewline}{\\}

\newcommand{\aap}{Astron. Astrophys.}
\newcommand{\pre}{Phys. Rev. E}
\newcommand{\apj}{Astrophys. J.}

\newcommand{\mnras}{Mon. Not. Roy. Astron. Soc.}
\newcommand{\solphys}{Sol. Phys.}
\newcommand{\physrep}{Phys. Rep.}
\newcommand{\EMF}{\mbox{\boldmath ${\cal E}$} {}}

\newcommand{\ccdot}{{\bm\cdot}}
\newcommand{\mathd}{\mathrm{d}}
\newcommand{\mathe}{\mathrm{e}}

\received{\today,~ $ $Revision: 1.20 $ $}

\title{Helicity--vorticity turbulent pumping of magnetic fields in the solar
convection zone}

\doi{10.1080/03091920xxxxxxxxx}
\issn{1029-0419} \issnp{0309-1929} \jvol{00} \jnum{00} \jyear{2013}
\author{V. V. PIPIN
\vspace{6pt}\\
Institute Solar-Terrestrial Physics, Irkutsk, Russia
}

\begin{document}

\maketitle
\begin{abstract}
We study the effect of turbulent drift of a large-scale magnetic
field that results from the interaction of helical convective
motions and differential rotation in the solar convection zone.
The principal direction of the drift corresponds to the direction of the
large-scale vorticity vector. Thus, the effect produces a latitudinal
transport of the large-scale magnetic field in the convective zone
wherever the angular velocity has a strong radial gradient.
The direction of the drift depends on the sign of helicity and it is
defined by the Parker-Yoshimura rule. The analytic calculations are
done within the framework of mean-field magnetohydrodynamics
using the minimal $\tau$-approximation. We estimate the
magnitude of the drift velocity and find that it can be several m/s
near the base of the solar convection zone. The implications of this
effect for the solar dynamo are illustrated on the basis of an
axisymmetric mean-field dynamo model with a subsurface shear layer.
We find that the helicity--vorticity pumping effect can have an influence
on the features of the sunspot time--latitude diagram, producing a
fast drift of the sunspot activity maximum at the rise phase of the
cycle and a slow drift at the decay phase of the cycle.
\end{abstract}

\section{Introduction}

It is believed that the evolution of the large-scale magnetic field of
the Sun is governed by the interplay between large-scale motions,
like differential rotation and meridional circulation, turbulent
convection flows and magnetic fields. One of the most important issues
in solar dynamo theory is related to the origin of the equatorial
drift of sunspot activity in the equatorial regions and, simultaneously
at high latitudes, the poleward drift of the location of large-scale unipolar
regions and quiet prominences.
\cite{par55} and \cite{yosh1975} suggested that the evolution of large-scale magnetic
activity of the Sun can be interpreted as dynamo waves propagating
perpendicular to the direction of shear from the differential rotation.
They found that the propagation can be considered as a diffusion process,
which follows the iso-rotation surfaces of angular velocity in
the solar convection zone. The direction of propagation can be modified
by meridional circulation, anisotropic diffusion and the effects
of turbulent pumping (see, e.g., \citealp{choud95,k02,2008A&A...485..267G}).
The latter induces an effective drift of the large-scale magnetic field
even though the mean flow of the turbulent medium may be zero. 

The turbulent pumping effects can be equally important both for
dynamos without meridional circulation and for the meridional
circulation-dominated dynamo regimes. For the latter case the velocity of
turbulent pumping has to be comparable to the meridional circulation
speed. It is known that an effect of this magnitude can be produced
by diamagnetic pumping and perhaps by so-called topological
pumping. Both effects produce pumping in the radial direction
and have not a direct impact on the latitudinal drift of the
large-scale magnetic field. 

\begin{figure}
\begin{centering}
\includegraphics[width=0.6\textwidth]{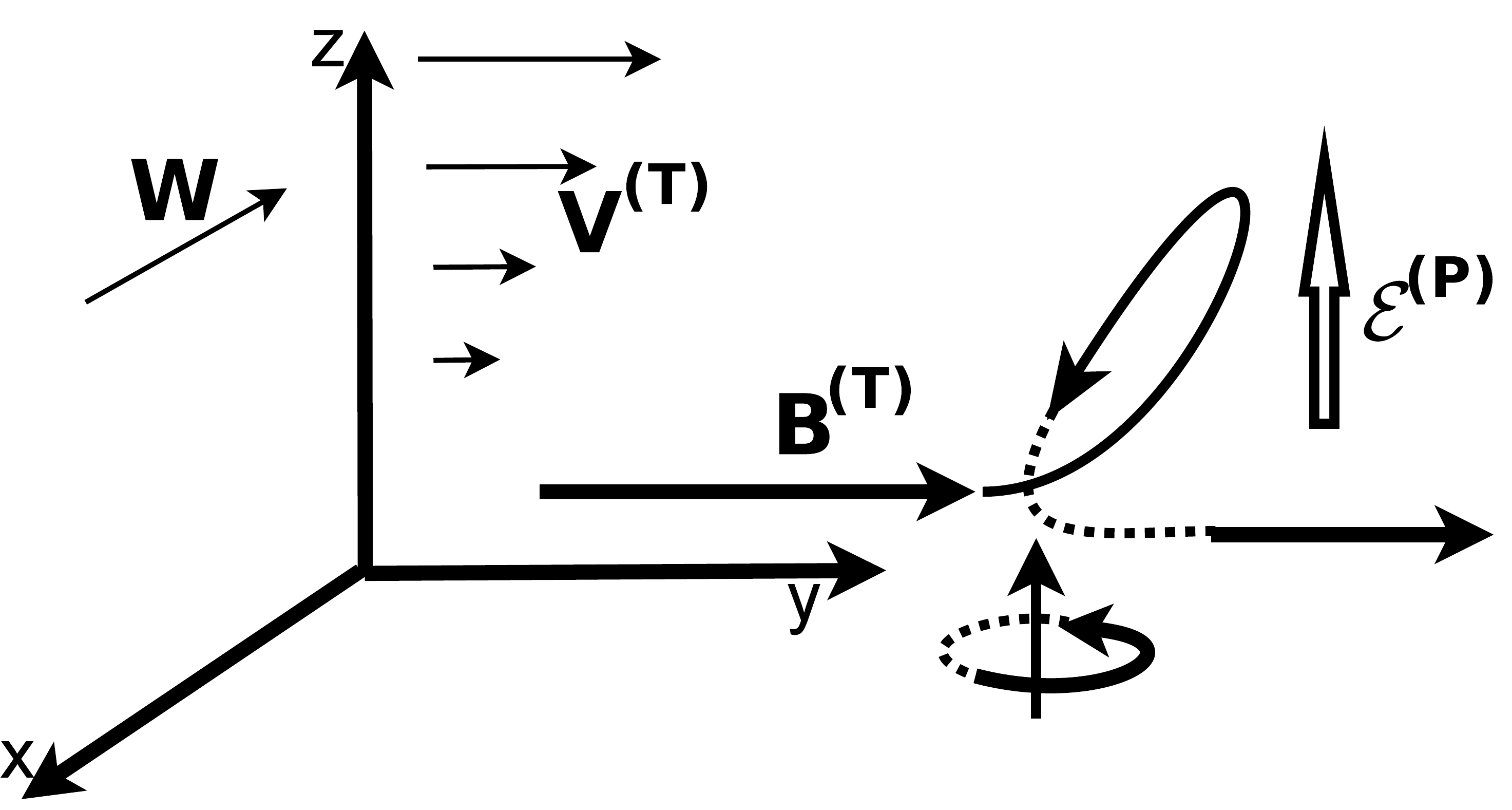}
\par\end{centering}
\caption{\label{fig:The-field-line}The field lines of the large-scale magnetic
field, $\bm{B}^{(T)}$, are transformed by the helical motions
to a twisted $\varOmega$-like shape. This loop is folded by the large-scale
shear, $\bm{V}^{(T)}$, into the direction of the background large-scale
magnetic field, $\bm{B}^{(T)}$. The induced electromotive force
has a component, $\EMF^{(P)}$, which is perpendicular to
the field $\bm{B}^{(T)}$. The resulting effect is identical to
the effective drift of the large-scale magnetic field along the $x$-axis,
in the direction opposite to the large-scale vorticity vector $\bm{W}=\nabla\times\bm{V}^{(T)}$,
i.e., $\EMF^{(P)}\sim-\bm{W}\times\bm{B}^{(T)}$.}
\end{figure}

Recently \citep{pi08Gafd,mitr2009AA,2010GApFD.104..167L}, it has been
found that the helical convective motions and the helical turbulent
magnetic fields interacting with large-scale magnetic fields and
differential rotation can produce effective pumping in the direction
of the large-scale vorticity vector. Thus, the effect produces a
latitudinal transport of the large-scale magnetic field in the convective
zone wherever the angular velocity has a strong radial gradient.
It is believed that these regions, namely the tachocline beneath
the solar convection zone and the subsurface shear layer,
are important for the solar dynamo. Figure \ref{fig:The-field-line}
illustrates the principal processes that induce the helicity--vorticity
pumping effect. It is suggested that this effect produces an
anisotropic drift of the large-scale magnetic field, which means that
the different components of the large-scale magnetic field drift in
different directions. Earlier work, e.g.\ by \citet{kit:1991} and \citet{2003PhRvE..67b6321K},
suggests that the effect of anisotropy in the transport of mean-field
is related to nonlinear effects of the global Coriolis force
on the convection. Also, nonlinear effects of the large-scale
magnetic field result in an anisotropy of turbulent pumping \citep{1996A&A...307..293K}.
It is noteworthy, that the helicity--vorticity effect produces an anisotropy
of the large-scale magnetic field drift already in the case of
slow rotation and a weak magnetic field. A comprehensive study
of the linear helicity--vorticity pumping effect for the case of weak
shear and slow rotation was given by \citet{garr2011} and their results
were extended by DNS with a more general test-field method \citet{bran2012AA}.

In this paper we analytically estimate the helicity--vorticity pumping
effect taking into account the Coriolis force due to global rotation.
The calculations were done within the framework of mean-field magnetohydrodynamics
using the minimal $\tau$-approximation.
The results are applied to mean field dynamo models, which are
used to examine this effect on the dynamo. The paper is structured
as follows. In the next section we briefly outline the basic equations
and assumptions, and consider the results of calculations. Next, we apply
the results to the solar dynamo. In Section 3 we summarize the
main results of the paper. The details of analytical calculations
are given in the Appendices A and B.

\section{Basic equations}

In the spirit of mean-field magnetohydrodynamics, we split the
physical quantities of the turbulent conducting fluid into mean
and fluctuating parts where the mean part is defined as an ensemble
average. One assumes the validity of the Reynolds
rules. The magnetic field $\bm{B}$ and the velocity $\bm{V}$
are decomposed as $\bm{B}=\overline{\bm{B}}+\bm{b}$ and
$\bm{V}=\overline{\bm{V}}+\bm{u}$, respectively. Hereafter,
we use small letters for the fluctuating parts and
capital letters with an overbar for mean fields. Angle brackets
are used for ensemble averages of products. We use the
two-scale approximation \citep{rob-saw,krarad80} and assume that
mean fields vary over much larger scales (both in time and
in space) than fluctuating fields. The average effect of MHD-turbulence
on the large-scale magnetic field (LSMF) evolution is described by the
mean-electromotive force (MEMF), 
$\EMF=\left\langle \bm{u\times b}\right\rangle$.
The governing equations for fluctuating magnetic field and velocity
are written in a rotating coordinate system as follows:
\begin{eqnarray}
\frac{\partial\bm{b}}{\partial t} & = & \nabla\times\left(\bm{u}\times\bm{\bar{B}}+\bm{\bar{V}}\times\bm{b}\right)+\eta\nabla^{2}\bm{b}+\mathfrak{G},\label{induc1}\\
\frac{\partial u_{i}}{\partial t}+2\left(\bm{\varOmega}\times\bm{u}\right)_{i} & = & -\nabla_{i}\left(p+\frac{\left(\bm{b\ccdot\bar{B}}\right)}{\mu}\right)+\nu\Delta u_{i}\label{navie1}\\
 & + & \frac{1}{\mu}\nabla_{j}\left(\bar{B}_{j}b_{i}+\bar{B}_{i}b_{j}\right)-\nabla_{j}\left(\bar{V}_{j}u_{i}+\bar{V}_{i}u_{j}\right)+f_{i}+\mathfrak{F}_{i},\nonumber 
\end{eqnarray}
where $\mathfrak{G},\mathfrak{F}$ stand for nonlinear contributions
to the fluctuating fields, $p$ is the fluctuating pressure, $\bm{\varOmega}$
is the angular velocity responsible for the Coriolis force, $\bm{\bar{V}}$
is mean flow which is a weakly variable in space, and $\bm{f}$ is the
random force driving the turbulence. Equations~(\ref{induc1})
and (\ref{navie1}) are used to compute the mean-electromotive force
$\EMF=\left\langle \bm{u\times b}\right\rangle $.
It was computed with the help of the equations for the second moments of
fluctuating velocity and magnetic fields using the double-scale Fourier
transformation and the minimal $\tau$-approximations and for a
given model of background turbulence. To simplify the estimation
of nonlinear effects due to global rotation, we use scale-independent
background turbulence spectra and correlation time. Details of the
calculations are given in Appendix A. In what follows we discuss
only those parts of the mean-electromotive force which are related to
shear and the pumping effect.

\subsection{Results}

The large-scale shear flow is described by the tensor $\overline{V}_{i,j}=\nabla\overline{V}_{i}$.
It can be decomposed into a sum of strain and vorticity tensors,
$\nabla_{j}\overline{V}_{i}={\displaystyle
  \frac{1}{2}\left(\overline{V}_{i,j}
+\overline{V}_{j,i}\right)-\frac{1}{2}\varepsilon_{ijp}\overline{W}_{p}}$,
where $\overline{\bm{W}}=\boldsymbol{\nabla}\times\overline{\bm{V}}$
is the large-scale vorticity vector. The joint effect of large-scale
shear, helical turbulent flows and magnetic fields can be expressed
by the following contributions to the mean-electromotive force
(omitting the $\alpha$-effect):
\begin{eqnarray}
\EMF^{(H)} & = & \left(\bm{\overline{W}}\times\bm{\overline{B}}\right)\left(f_{2}^{(\gamma)}h_{\mathcal{C}}+f_{1}^{(\gamma)}h_{\mathcal{K}}\right)\tau_{c}^{2}+\bm{\tilde{V}}\left(\bm{B}\right)\left(f_{4}^{(\gamma)}h_{\mathcal{C}}+f_{3}^{(\gamma)}h_{\mathcal{K}}\right)\tau_{c}^{2}\label{eq:EMFP}\\
 & + & \bm{e}\left[\left(\bm{e}\times\bm{\overline{W}}\right)\ccdot\bm{\overline{B}}\right]\left(f_{6}^{(\gamma)}h_{\mathcal{C}}+f_{5}^{(\gamma)}h_{\mathcal{K}}\right)\tau_{c}^{2}+\left(\bm{e}\times\bm{\overline{W}}\right)\left(\bm{e}\ccdot\bm{\overline{B}}\right)\left(f_{8}^{(\gamma)}h_{\mathcal{C}}+f_{7}^{(\gamma)}h_{\mathcal{K}}\right)\tau_{c}^{2},\nonumber
\end{eqnarray}
where $\bm{\tilde{V}}\left(\bm{B}\right)={\displaystyle \frac{\overline{B}_{j}}{2}\left(\overline{V}_{i,j}+\overline{V}_{j,i}\right)}$,
${\displaystyle \bm{e}=\frac{\boldsymbol{\varOmega}}{\left|\boldsymbol{\varOmega}\right|}}$
is the unit vector along the rotation axis, $\tau_{c}$ is the typical
relaxation time of turbulent flows and magnetic fields, $h_{\mathcal{K}}^{\left(0\right)}=\left\langle \bm{u^{\left(0\right)}\ccdot\nabla\times u^{\left(0\right)}}\right\rangle $
and $h_{\mathcal{C}}^{\left(0\right)}={\displaystyle \frac{\left\langle \bm{b^{\left(0\right)}\ccdot\nabla\times b^{\left(0\right)}}\right\rangle }{\mu\rho}}$
are kinetic and current helicity of the background turbulence.
These parameters are assumed to be known in advance. Functions $f_{n}^{(\gamma)}(\varOmega^{*})$
are given in Appendix B, they depend on the Coriolis number $\varOmega^{*}=2\varOmega_{0}\tau_{c}$
and describe the nonlinear effect due the Coriolis force, and $\varOmega_{0}$
is the global rotation rate. 

\begin{figure}
\begin{centering}
\includegraphics[width=0.45\textwidth]{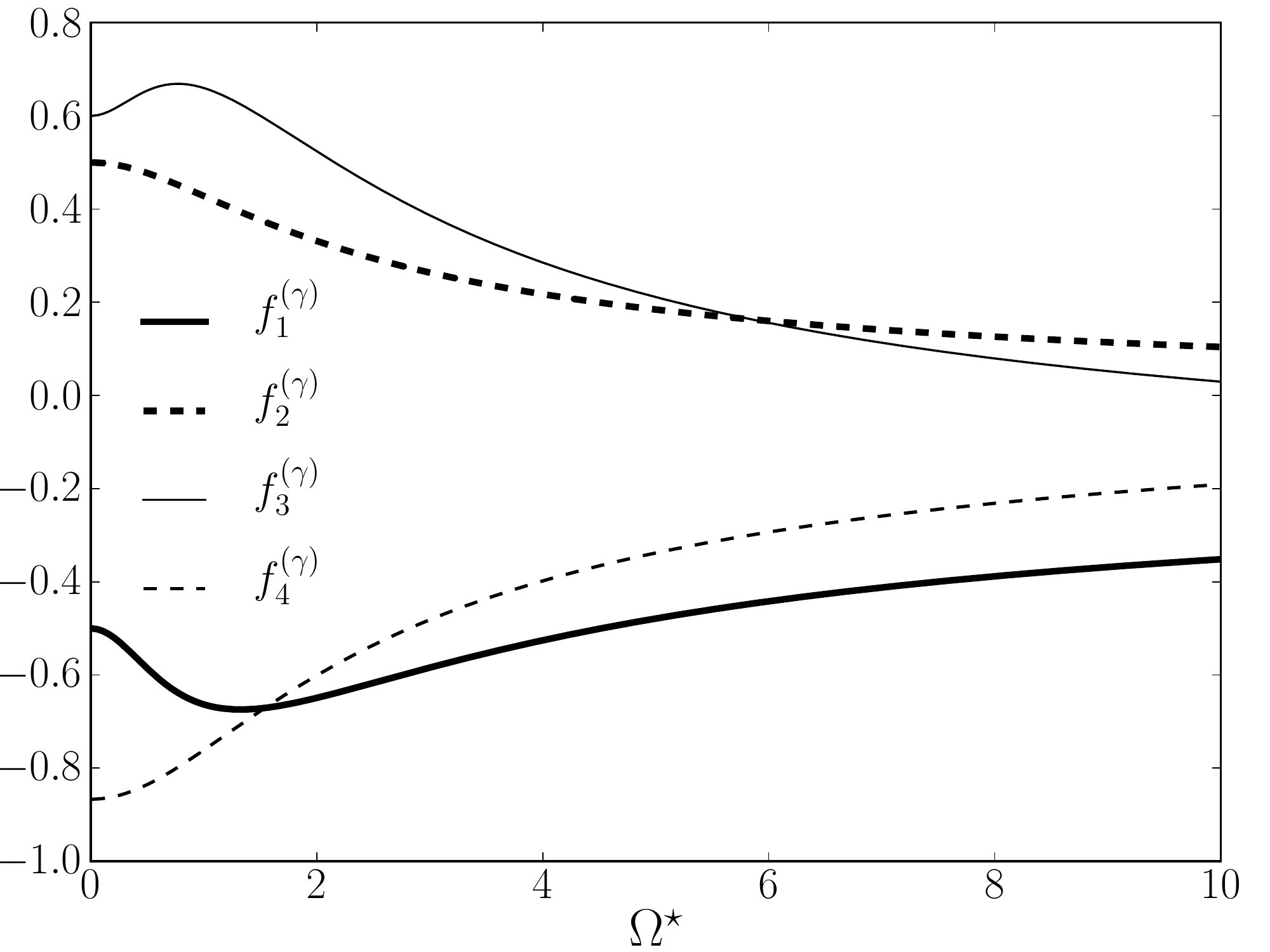}\includegraphics[width=0.45\textwidth]{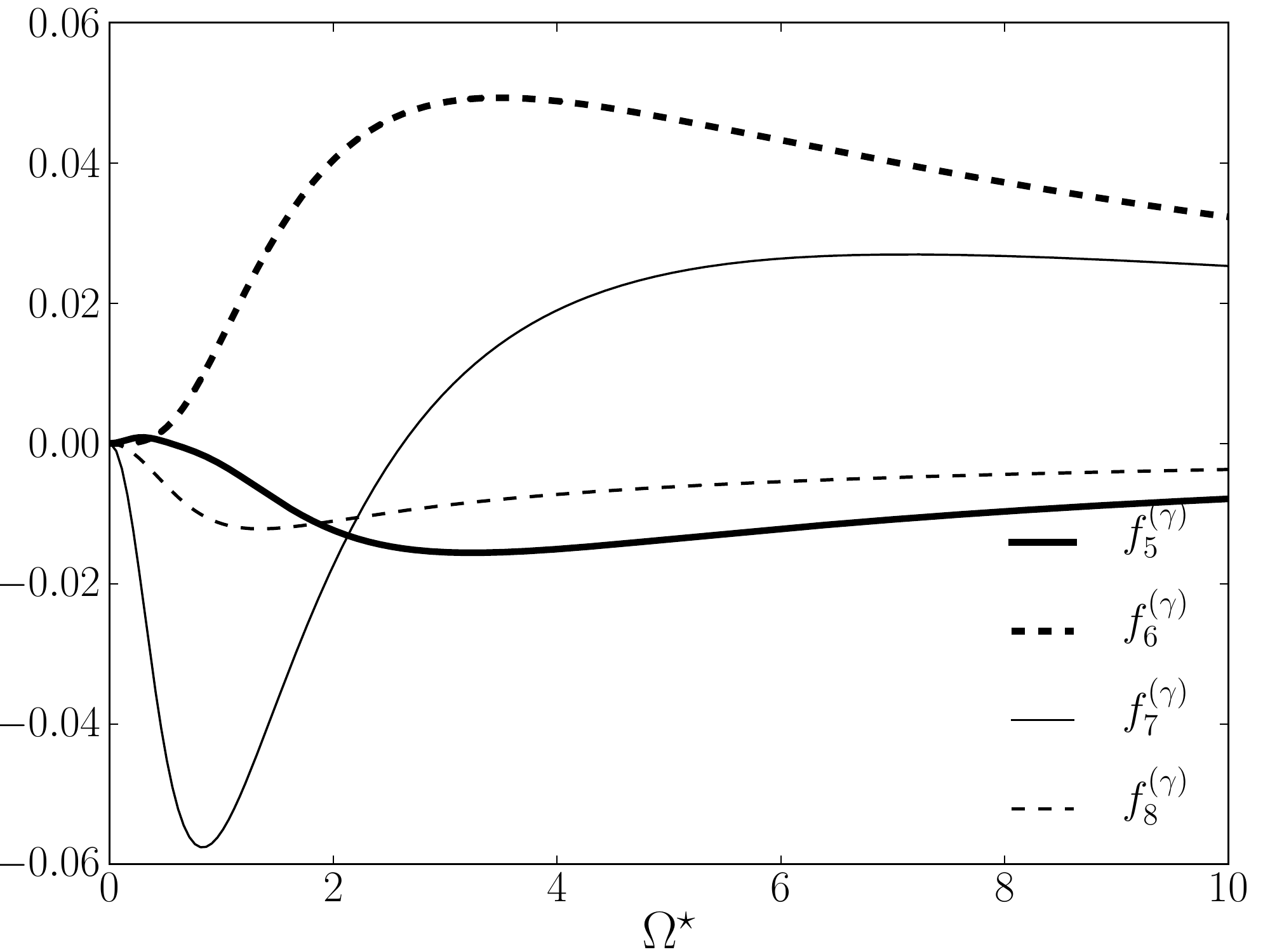}
\par\end{centering}

\caption{\label{fig:Fun-dep}The dependence of the pumping effects on the Coriolis
number. Solid lines show contributions from kinetic helicity and
dashed lines the same for current helicity.}
\end{figure}

For slow rotation, $\varOmega^{*}\ll1$,
we perform a Taylor expansion of $f_{n}^{(\gamma)}(\varOmega^{*})$ and
obtain
\begin{eqnarray}
\EMF^{(H)} & = & \frac{\tau_{c}^{2}}{2}\left(\bm{\overline{W}}\times\bm{\overline{B}}\right)\left(h_{\mathcal{C}}-h_{\mathcal{K}}\right)+\frac{\tau_{c}^{2}}{5}\bm{\tilde{V}}\left(\bm{B}\right)\left(3h_{\mathcal{K}}-\frac{13}{3}h_{\mathcal{C}}\right).\label{eq:EMFT}
\end{eqnarray}
The coefficients in the kinetic part of Eq.~(\ref{eq:EMFT}) are
two times larger than those found by \citet{garr2011}. This difference
results from our assumption that the background turbulence spectra
and the correlation time are scale-independent. The results for the magnetic
part are in agreement with our earlier findings (see \citealp{pi08Gafd}).
The first term in Eq.~(\ref{eq:EMFT}) describes turbulent pumping
with an effective velocity 
${\displaystyle \frac{\tau_{c}^{2}\bm{\overline{W}}}{2}\left(h_{\mathcal{C}}-h_{\mathcal{K}}\right)}$
and the second term describes anisotropic turbulent pumping. Its
structure depends on the geometry of the shear flow.
For large Coriolis
numbers, $\varOmega^{*}\gg1$, only the kinetic helicity contributions
survive:
\begin{equation}
\EMF^{(H)}=-\frac{\tau_{c}^{2}}{6}\left(\bm{\overline{W}}\times\bm{\overline{B}}\right)h_{\mathcal{K}}+\frac{\tau_{c}^{2}}{5}\bm{\tilde{V}}\left(\bm{B}\right)h_{\mathcal{K}}.\label{eq:EMFTI}
\end{equation}
 Figure \ref{fig:Fun-dep} show the dependence of the pumping effects
on the Coriolis number. We observe that for the terms $\left(\bm{\overline{W}}\times\bm{\overline{B}}\right)$
and $\bm{\tilde{V}}\left(\bm{B}\right)$ the effects of
kinetic helicity are non-monotonic and have a maximum at $\varOmega^{*}\approx1$.
The effects of current helicity for these terms are monotonically
quenched with increasing values of $\varOmega^{*}$. The additional contributions
in Eq.~(\ref{eq:EMFP}) are rather small in comparison with the main
terms. Thus, we can conclude that the first line in Eq.~(\ref{eq:EMFP}) describes the leading
effect of pumping due to the helicity of turbulent flows and magnetic
field. Below, we drop the contributions from the second line in Eq.~(\ref{eq:EMFP})
from our analysis.

\subsection{Helicity--vorticity pumping in the solar convection zone}

\subsubsection{The dynamo model}

To estimate the impact of this pumping effect on the dynamo we consider
the example of a dynamo model which takes into account contributions
of the mean electromotive force given by Eq.~(\ref{eq:EMFP}). The
dynamo model employed in this paper has been described in detail
by \cite{pk11,pk11mf}. This type of dynamo was proposed originally
by \citet{2005ApJ...625..539B}. The reader may find the discussion
for different types of mean-field dynamos in \cite{2005PhR...417....1B}
and \cite{2007sota.conf..319T}.

We study the standard mean-field induction equation in a perfectly conducting
medium: 
\begin{equation}
\frac{\partial\overline{\bm{B}}}{\partial t}=\boldsymbol{\nabla}\times\left(\EMF+\overline{\bm{U}}\times\overline{\bm{B}}\right),\label{eq:dyn}
\end{equation}
where $\EMF=\overline{\bm{u\times b}}$ is
the mean electromotive force, with $\bm{u,\, b}$ being
fluctuating velocity and magnetic field, respectively, $\overline{\bm{U}}$
is the mean velocity (differential rotation and meridional circulation),
and the axisymmetric magnetic field is:
\[
\overline{\bm{B}}=\bm{e}_{\phi}B+\nabla\times\frac{A\bm{e}_{\phi}}{r\sin\theta},
\]
where $\theta$ is the polar angle. The expression for the mean electromotive
force $\EMF$ is given by \citet{pi08Gafd}.
It is expressed as follows: 
\begin{equation}
\mathcal{E}_{i}=\left(\alpha_{ij}+\gamma_{ij}^{(\varLambda)}\right)\overline{B}-\eta_{ijk}\nabla_{j}\overline{B}_{k}+\mathcal{E}_{i}^{(H)}.\label{eq:EMF-1}
\end{equation}
The new addition due to helicity and mean vorticity effects is
marked by $\EMF^{H}$. The tensor $\alpha_{ij}$ represents
the $\alpha$-effect. It includes hydrodynamic and magnetic helicity
contributions, 
\begin{align}
\alpha_{ij} & =C_{\alpha}\sin^{2}\theta\alpha_{ij}^{(H)}+\alpha_{ij}^{(M)},\label{alp2d}\\
\alpha_{ij}^{(H)} & =\delta_{ij}\left\{ 3\eta_{T}\left(f_{10}^{(a)}\left(\bm{e}\ccdot\boldsymbol{\varLambda}^{(\rho)}\right)+f_{11}^{(a)}\left(\bm{e}\ccdot\boldsymbol{\varLambda}^{(u)}\right)\right)\right\} +\\
 & +e_{i}e_{j}\left\{ 3\eta_{T}\left(f_{5}^{(a)}\left(\bm{e}\ccdot\boldsymbol{\varLambda}^{(\rho)}\right)+f_{4}^{(a)}\left(\bm{e}\ccdot\boldsymbol{\varLambda}^{(u)}\right)\right)\right\} +\\
 & 3\eta_{T}\left\{ \left(e_{i}\varLambda_{j}^{(\rho)}+e_{j}\varLambda_{i}^{(\rho)}\right)f_{6}^{(a)}+\left(e_{i}\varLambda_{j}^{(u)}+e_{j}\varLambda_{i}^{(u)}\right)f_{8}^{(a)}\right\} ,
\end{align}
where the hydrodynamic part of the $\alpha$-effect is defined by
$\alpha_{ij}^{(H)}$, $\bm{\boldsymbol{\varLambda}}^{(\rho)}=\boldsymbol{\nabla}\log\overline{\rho}$
quantifies the density stratification, $\bm{\boldsymbol{\varLambda}}^{(u)}=\boldsymbol{\nabla}\log\left(\eta_{T}^{(0)}\right)$
quantifies the turbulent diffusivity variation, and $\bm{e}=\boldsymbol{\varOmega}/\left|\varOmega\right|$
is a unit vector along the axis of rotation.
The turbulent pumping, $\gamma_{ij}^{(\varLambda)}$, depends
on mean density and turbulent diffusivity
stratification, and on the Coriolis number $\varOmega^{*}=2\tau_{c}\varOmega_{0}$
where $\tau_{c}$ is the typical convective turnover time and $\varOmega_{0}$
is the global angular velocity. Following the results of \cite{pi08Gafd}, $\gamma_{ij}^{(\varLambda)}$
is expressed as follows: 
\begin{align}
\gamma_{ij}^{(\varLambda)} & =3\eta_{T}\left\{ f_{3}^{(a)}\varLambda_{n}^{(\rho)}+f_{1}^{(a)}\left(\bm{e}\ccdot\boldsymbol{\varLambda}^{(\rho)}\right)e_{n}\right\} \varepsilon_{inj}-3\eta_{T}f_{1}^{(a)}e_{j}\varepsilon_{inm}e_{n}\varLambda_{m}^{(\rho)},\label{eq:pump}\\
 & -3\eta_{T}\left(\varepsilon-1\right)\left\{ f_{2}^{(a)}\varLambda_{n}^{(u)}+f_{1}^{(a)}\left(\bm{e}\ccdot\boldsymbol{\varLambda}^{(u)}\right)e_{n}\right\} \varepsilon_{inj}.
\end{align}
The effect of turbulent diffusivity, which is anisotropic due to the
Coriolis force, is given by: 
\begin{equation}
\eta_{ijk}=3\eta_{T}\left\{ \left(2f_{1}^{(a)}-f_{2}^{(d)}\right)\varepsilon_{ijk}-2f_{1}^{(a)}e_{i}e_{n}\varepsilon_{njk}+\varepsilon C_{\omega}f_{4}^{(d)}e_{j}\delta_{ik}\right\}.\label{eq:diff}
\end{equation}
The last term in Eq.~(\ref{eq:diff}) describes R\"adler's $\boldsymbol{\varOmega}\times\bm{J}$
effect. The functions $f_{\{1-11\}}^{(a,d)}$ depend on the Coriolis number.
They can be found in \cite{pi08Gafd}; see also \cite{pk11} or \cite{ps11}). In the
model, the parameter $\varepsilon={\displaystyle \frac{\overline{\bm{b}^{2}}}{\mu_{0}\overline{\rho}\overline{\bm{u}^{2}}}}$,
which measures the ratio between magnetic and kinetic energies
of the fluctuations in the background turbulence, is assumed to be equal to
1. This corresponds to perfect energy equipartition. The $\varepsilon$
contribution in the second line of Eq.~(\ref{eq:pump}) describes the
paramagnetic effect \citep{2003PhRvE..67b6321K}. In the state of
perfect energy equipartition the effect of diamagnetic pumping is
compensated by the paramagnetic effect. We can, formally, skip
the second line in Eq.~(\ref{eq:pump}) from our consideration if
$\varepsilon=1$. To compare the magnitude of the helicity--vorticity
pumping effect with the diamagnetic effect we will show results for
the pumping velocity distribution with $\varepsilon=0$.

The contribution of small-scale magnetic helicity $\overline{\chi}=\overline{\bm{a\ccdot}\bm{b}}$
($\bm{a}$ is the fluctuating vector-potential of the magnetic field)
to the $\alpha$-effect is defined as
\begin{equation}
\alpha_{ij}^{(M)}=2f_{2}^{(a)}\delta_{ij}\frac{\overline{\chi}\tau_{c}}{\mu_{0}\overline{\rho}\ell^{2}}-2f_{1}^{(a)}e_{i}e_{j}\frac{\overline{\chi}\tau_{c}}{\mu_{0}\overline{\rho}\ell^{2}}.\label{alpM}
\end{equation}
The nonlinear feedback of the large-scale magnetic field to the $\alpha$-effect
is described by
a dynamical quenching due
to the constraint of magnetic helicity conservation. The magnetic helicity,
$\overline{\chi}$ , subject to a conservation law, is described by
the following equation \citep{kle-rog99,sub-bra:04}: 
\begin{eqnarray}
\frac{\partial\overline{\chi}}{\partial t} & = & -2\left(\EMF\ccdot\overline{\bm{B}}\right)-\frac{\overline{\chi}}{R_{\chi}\tau_{c}}+\boldsymbol{\nabla}\ccdot\left(\eta_{\chi}\boldsymbol{\nabla}\bar{\chi}\right),\label{eq:hel}
\end{eqnarray}
where $\tau_{c}$ is a typical convective turnover time. The parameter
$R_{\chi}$ controls the helicity dissipation rate without specifying
the nature of the loss. The turnover time $\tau_{c}$ decreases from about 2 months
at the bottom of the integration domain, which is located at $0.71R_{\odot}$,
to several hours at the top boundary located at $0.99R_{\odot}$.
It seems reasonable that the helicity dissipation is most efficient
near the surface. The last term in Eq.~(\ref{eq:hel}) describes
a turbulent diffusive flux of magnetic helicity \citep{mitra10}.

\begin{figure}
\begin{centering}
\includegraphics[width=0.3\textwidth]{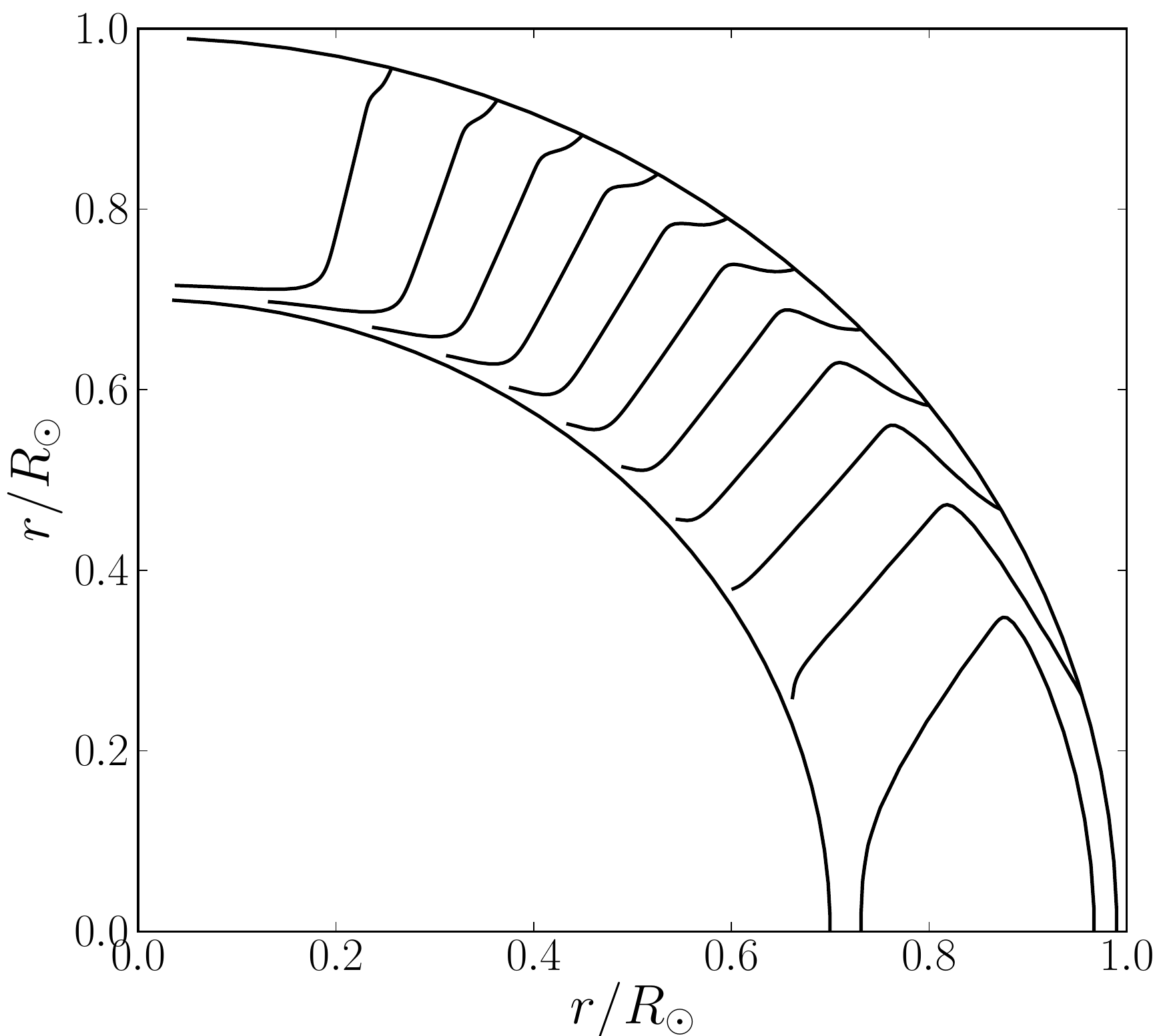}\includegraphics[width=0.35\textwidth]{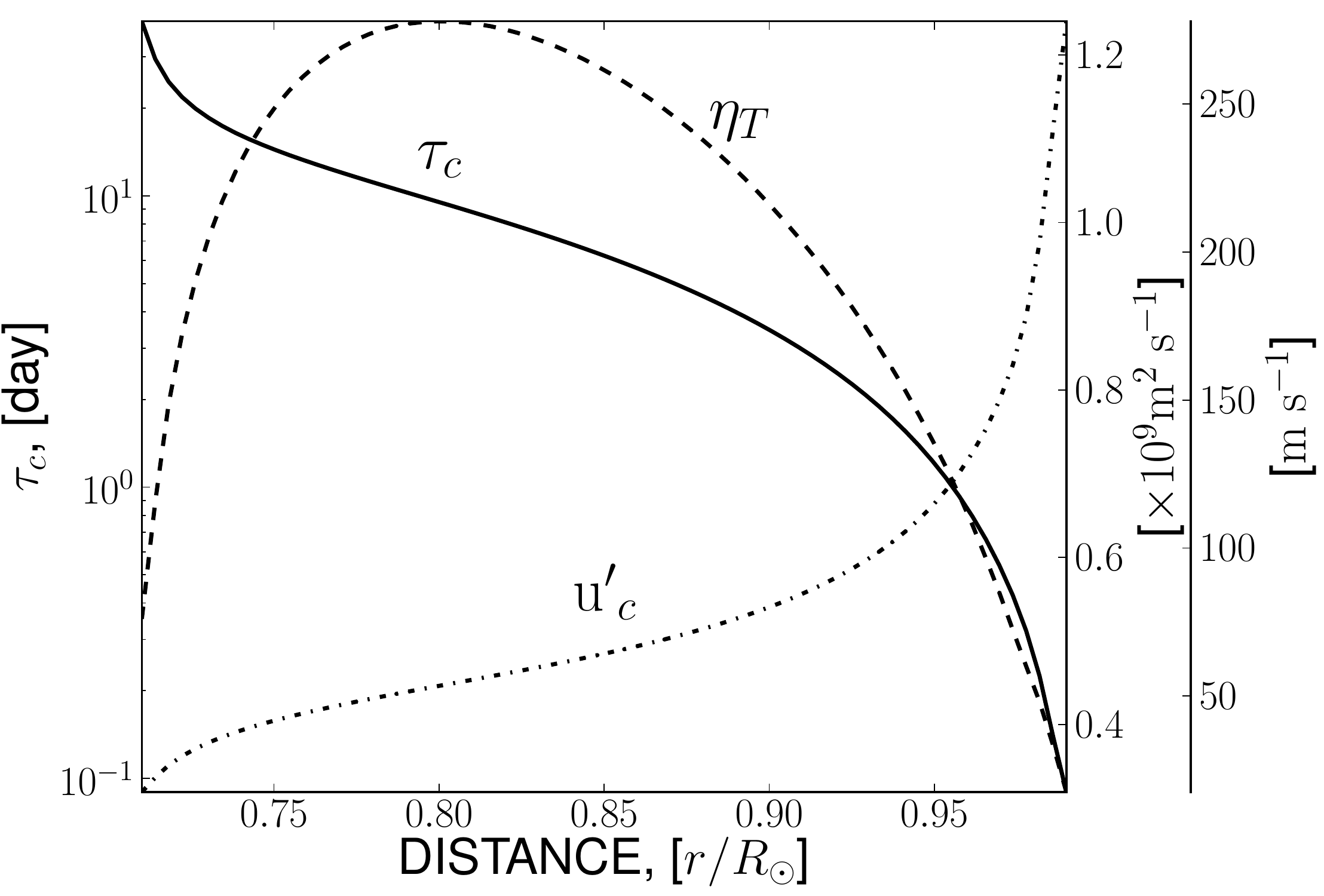}\includegraphics[width=0.35\textwidth]{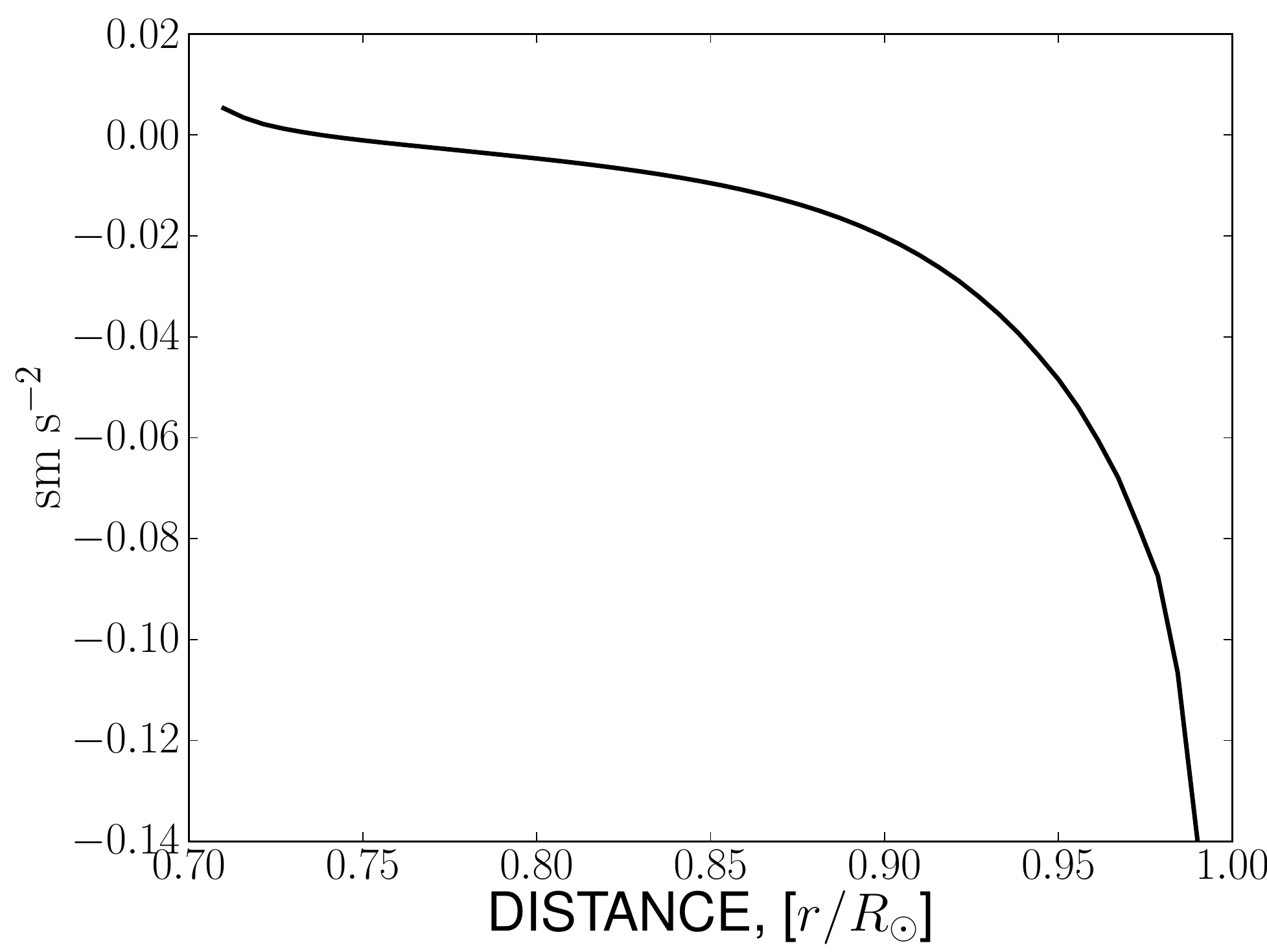}
\par\end{centering}

\begin{centering}
\includegraphics[width=0.45\textwidth]{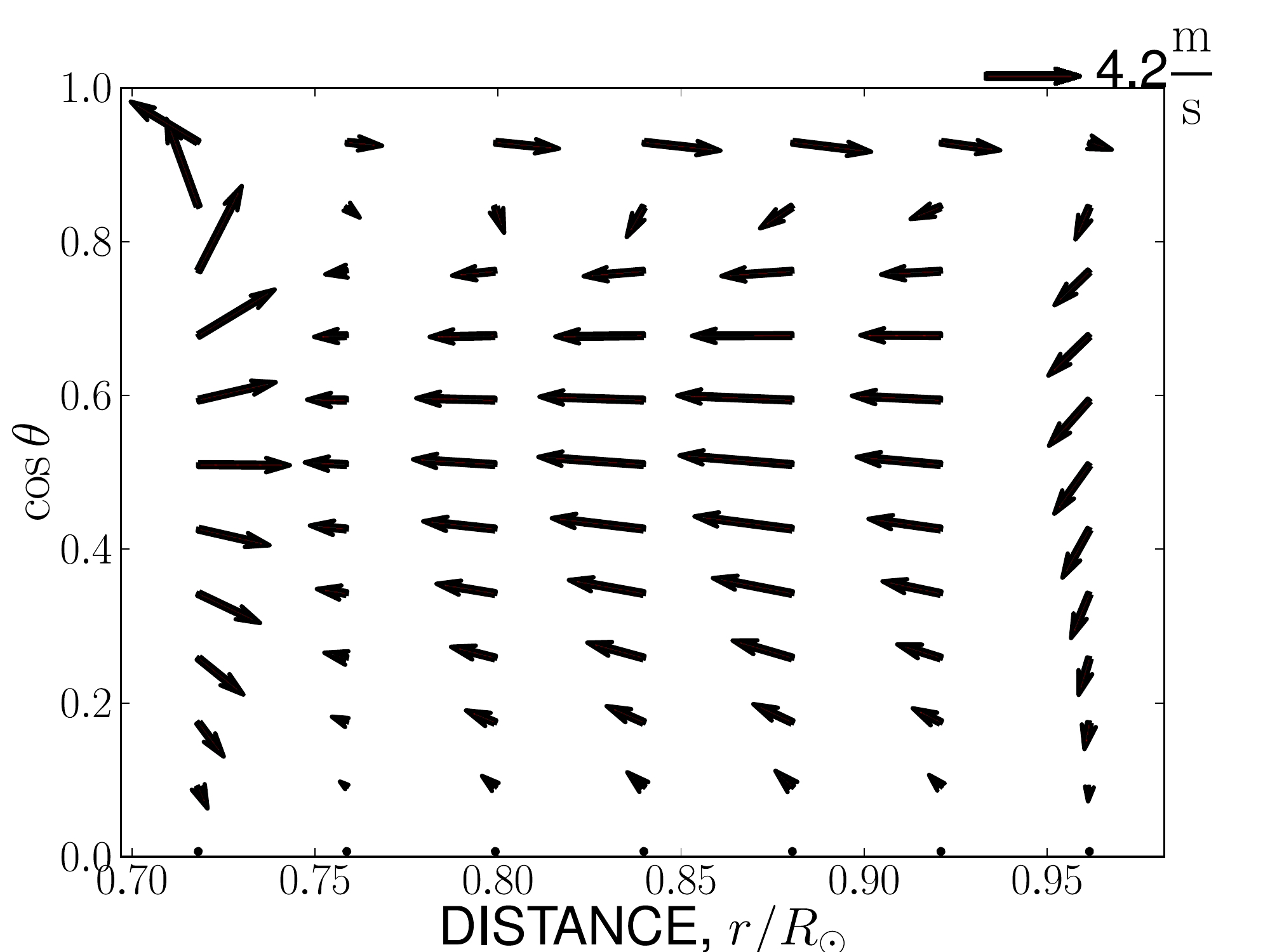}\includegraphics[width=0.45\textwidth]{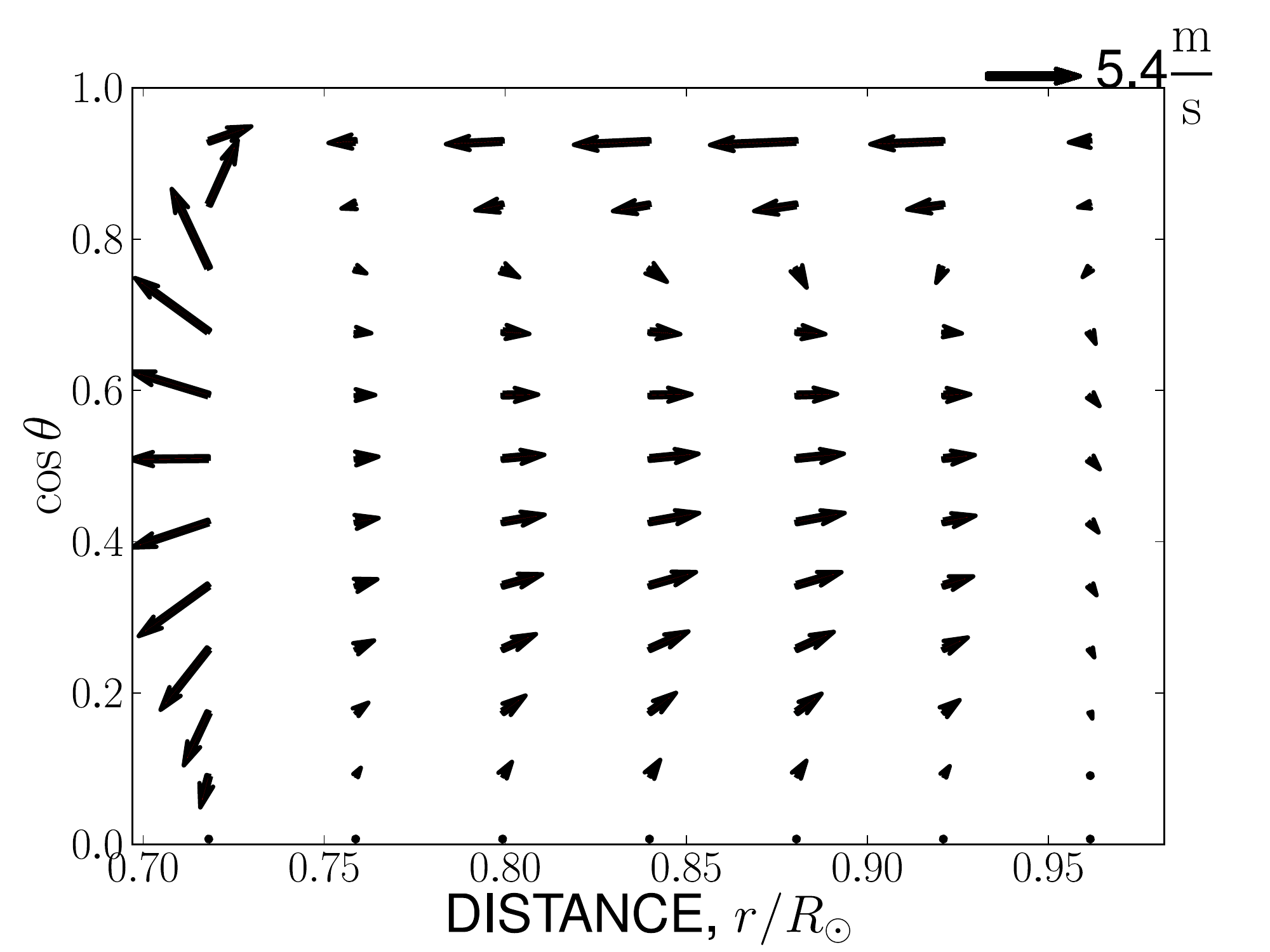}
\par\end{centering}

\caption{\label{fig:CZ}Distributions of the angular velocity and the turbulent
parameters, and the kinetic helicity inside the solar convection zone. The bottom panel shows
the patterns of the pumping velocity fields for the toroidal magnetic
field(left) and for the poloidal field(right). They were computed
on the basis of Eqs.~(\ref{eq:erH},\ref{eq:etH},\ref{eq:ephH}).}

\end{figure}

We use the solar convection zone model computed by \citet{stix:02},
in which the mixing-length is defined as $\ell=\alpha_{\rm MLT}\left|\varLambda^{(p)}\right|^{-1}$,
where $\bm{\boldsymbol{\varLambda}}^{(p)}=\boldsymbol{\nabla}\log\overline{p}\,$
quantifies the pressure variation, and $\alpha_{\rm MLT}=2$. The turbulent
diffusivity is parameterized in the form, $\eta_{T}=C_{\eta}\eta_{T}^{(0)}$,
where $\eta_{T}^{(0)}={\displaystyle \frac{u'\ell}{3}}$ is the characteristic
mixing-length turbulent diffusivity, $\ell$ is the typical correlation
length of the turbulence, and $C_{\eta}$ is a constant to control
the efficiency of large-scale magnetic field dragging by the turbulent
flow. Currently, this parameter cannot be introduced in the mean-field
theory in a consistent way. In this paper we use $C_{\eta}=0.05$.
The differential rotation profile, $\varOmega=\varOmega_{0}f_{\varOmega}\left(x,\mu\right)$
(shown in Fig.\ref{fig:CZ}a) is a slightly modified version of the
analytic approximation proposed by \citet{antia98}: 
\begin{eqnarray}
f_{\varOmega}\left(x,\mu\right) & = & \frac{1}{\varOmega_{0}}\left[\varOmega_{0}+55\left(x-0.7\right)\phi\left(x,x_{0}\right)\phi\left(-x,-0.96\right)\right.\label{eq:rotBA}\\
 & - & \left.200\left(x-0.95\right)\phi\left(x,0.96\right)\right)\nonumber \\
 & + & \left(21P_{3}\left(\mu\right)+3P_{5}\left(\mu\right)\right]\left(\frac{\mu^{2}}{j_{p}\left(x\right)}+\frac{1-\mu^{2}}{j_{e}\left(x\right)}\right)/\varOmega_{0},\nonumber \\
j_{p} & = & \frac{1}{1+\exp\left(\frac{0.709-x}{0.02}\right)},\,\, j_{e}=\frac{1}{1+\exp\left(\frac{0.692-x}{0.01}\right)},\nonumber
\end{eqnarray}
where $\varOmega_{0}=2.87\cdot10^{-6}s^{-1}$ is the equatorial angular
velocity of the Sun at the surface, $x=r/R_{\odot}$, $\phi\left(x,x_{0}\right)=0.5\left[1+\tanh\left[100(x-x_{0})\right]\right]$,
and $x_{0}=0.71$.

\subsubsection{Pumping effects in the solar convection zone}

The components of the strain tensor $\bm{\tilde{V}}$ in a spherical
coordinate system are given by the matrix: 
\[
\begin{array}{lll}
\bm{\tilde{V}} & = & \left(\begin{array}{ccc}
0 & 0 & \tilde{V}_{\left(r,\varphi\right)}\\
0 & 0 & \tilde{V}_{\left(\theta,\varphi\right)}\\
\tilde{V}_{\left(r,\varphi\right)} & \tilde{V}_{\left(\theta,\varphi\right)} & 0
\end{array}\right)\end{array},
\]
where we take into account only the azimuthal component of the large-scale
flow, $\tilde{V}_{\left(r,\varphi\right)}=r\sin\theta\partial_{r}\varOmega\left(r,\theta\right)$,
\ $\tilde{V}_{\left(\theta,\varphi\right)}=\sin\theta\partial_{\theta}\varOmega\left(r,\theta\right)$,
so $\bm{\hat{V}}\left(\bm{B}\right)=\left(B\tilde{V}_{\left(r,\varphi\right)},B\tilde{V}_{\left(\theta,\varphi\right)},\bm{B}_{i}^{p}\tilde{V}_{\left(i,\varphi\right)}\right)$.
Substituting this into Eq.~(\ref{eq:EMFP}) we find the components
of the mean-electromotive force for the helicity--vorticity pumping
effect, 
\begin{eqnarray}
\mathcal{E}_{r}^{(H)} & \!\!=\!\! & \frac{\varOmega^{\ast}\tau_{c}}{2}\sin\theta\left\{ \!\left[\! h_{\mathcal{K}}\left(f_{3}^{(\gamma)}-f_{1}^{(\gamma)}\right)+h_{\mathcal{C}}\left(f_{4}^{(\gamma)}-f_{2}^{(\gamma)}\right)\right]x\frac{\partial\tilde{\varOmega}}{\partial x}\!-\!2\left(\tilde{\varOmega}-1\right)\!\!\left[h_{\mathcal{K}}f_{1}^{(\gamma)}+h_{C}f_{2}^{(\gamma)}\right]\!\!\right\} \!\! B,\label{eq:erH}\\
\mathcal{E}_{\theta}^{(H)} & \!\!=\!\! & \frac{\varOmega^{\ast}\tau_{c}}{2}\left\{ \!\sin^{2}\theta\!\left[\! h_{\mathcal{K}}\left(f_{3}^{(\gamma)}-f_{1}^{(\gamma)}\right)\!+\! h_{\mathcal{C}}\left(f_{4}^{(\gamma)}-f_{2}^{(\gamma)}\right)\!\right]\!\!\frac{\partial\tilde{\varOmega}}{\partial\mu}-2\mu\left(\tilde{\varOmega}-1\right)\!\!\left[h_{\mathcal{K}}f_{1}^{(\gamma)}+h_{C}f_{2}^{(\gamma)}\right]\right\} \!\! B,\label{eq:etH}\\
\mathcal{E}_{\phi}^{(H)} & = & -\frac{\varOmega^{\ast}\tau_{c}}{2}\frac{\sin\theta}{x}\left[h_{\mathcal{K}}\left(f_{3}^{(\gamma)}+f_{1}^{(\gamma)}\right)+h_{\mathcal{C}}\left(f_{4}^{(\gamma)}+f_{2}^{(\gamma)}\right)\right]\frac{\partial\left(\tilde{\varOmega},A\right)}{\partial\left(x,\mu\right)}\label{eq:ephH}\\
 &  & -\frac{\left(\tilde{\varOmega}-1\right)\varOmega^{\ast}\tau_{c}}{x\sin\theta}\left[h_{\mathcal{K}}f_{1}^{(\gamma)}+h_{C}f_{2}^{(\gamma)}\right]\left(\mu\frac{\partial A}{\partial x}+\frac{\sin^{2}\theta}{x}\frac{\partial A}{\partial\mu}\right),\nonumber 
\end{eqnarray}
where $h_{\mathcal{C}}=C_{\mathcal{C}}{\displaystyle \frac{\overline{\chi}}{\mu_{0}\overline{\rho}\ell^{2}}}$.
It remains to define the kinetic helicity distribution. We use
a formula proposed in our earlier study (see \citealt{kps:06}),
\begin{align*}
h_{\mathcal{K}} & ={\displaystyle C_{\eta}C_{\mathcal{K}}}\frac{\overline{u^{(0)2}}}{2}\frac{\partial}{\partial r}\log\left(\overline{\rho}\sqrt{\overline{u^{(0)2}}}\right)F_{1}\cos\theta,
\end{align*}
where $F_{1}\left(\varOmega^{*}\right)$ was defined in the above cited paper.
The radial profile of ${\displaystyle \frac{h_{\mathcal{K}}}{\cos\theta}}$
is shown in Figure \ref{fig:CZ}. The radial profile of kinetic
helicity is shown in Figure 3a of the above cited paper. The parameters
$C_{\mathcal{K},\mathcal{C}}$ are introduced to switch on/off the
pumping effects in the model. 

\begin{figure}
\begin{centering}
\includegraphics[width=0.45\textwidth]{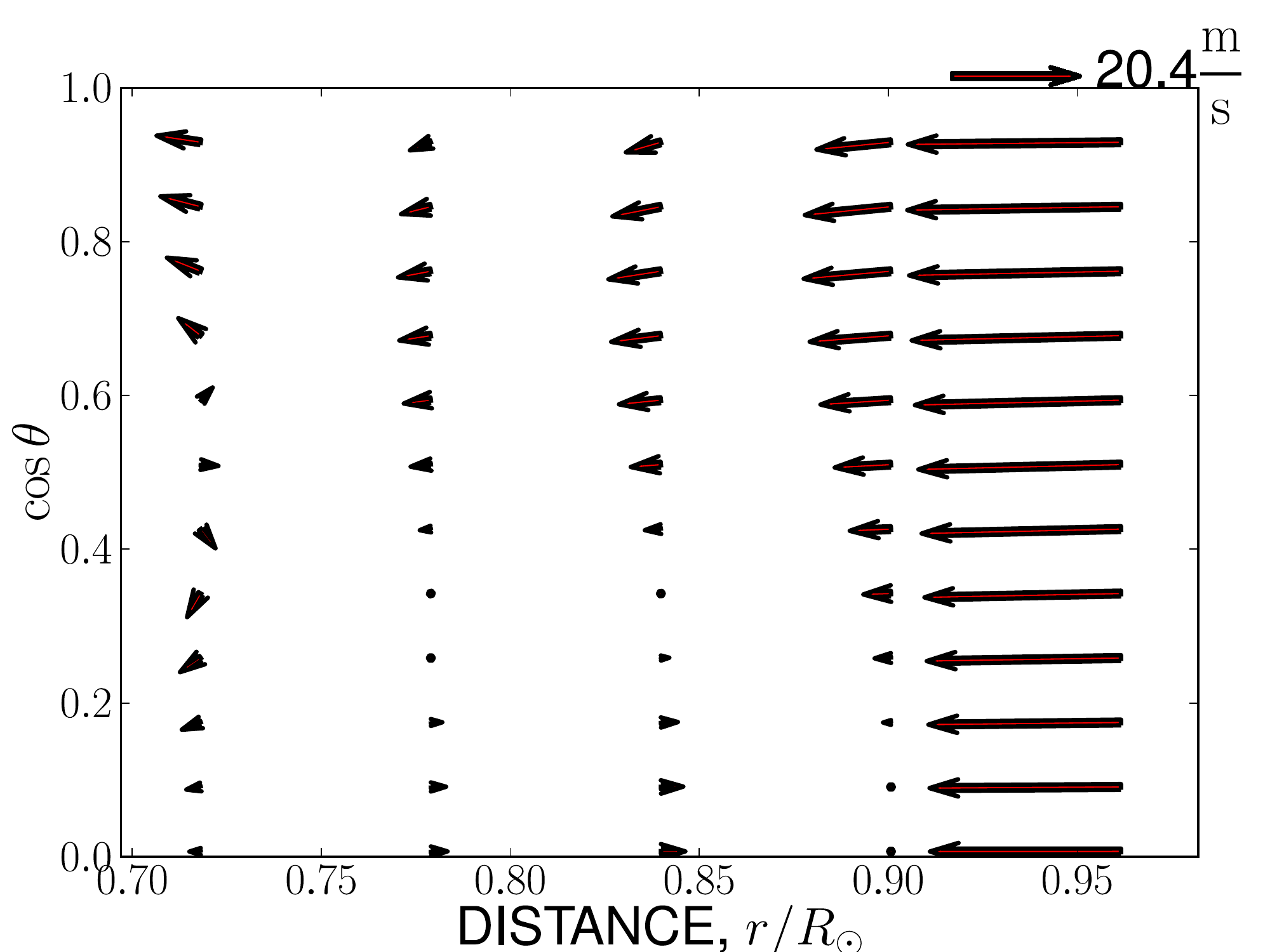}\includegraphics[width=0.45\textwidth]{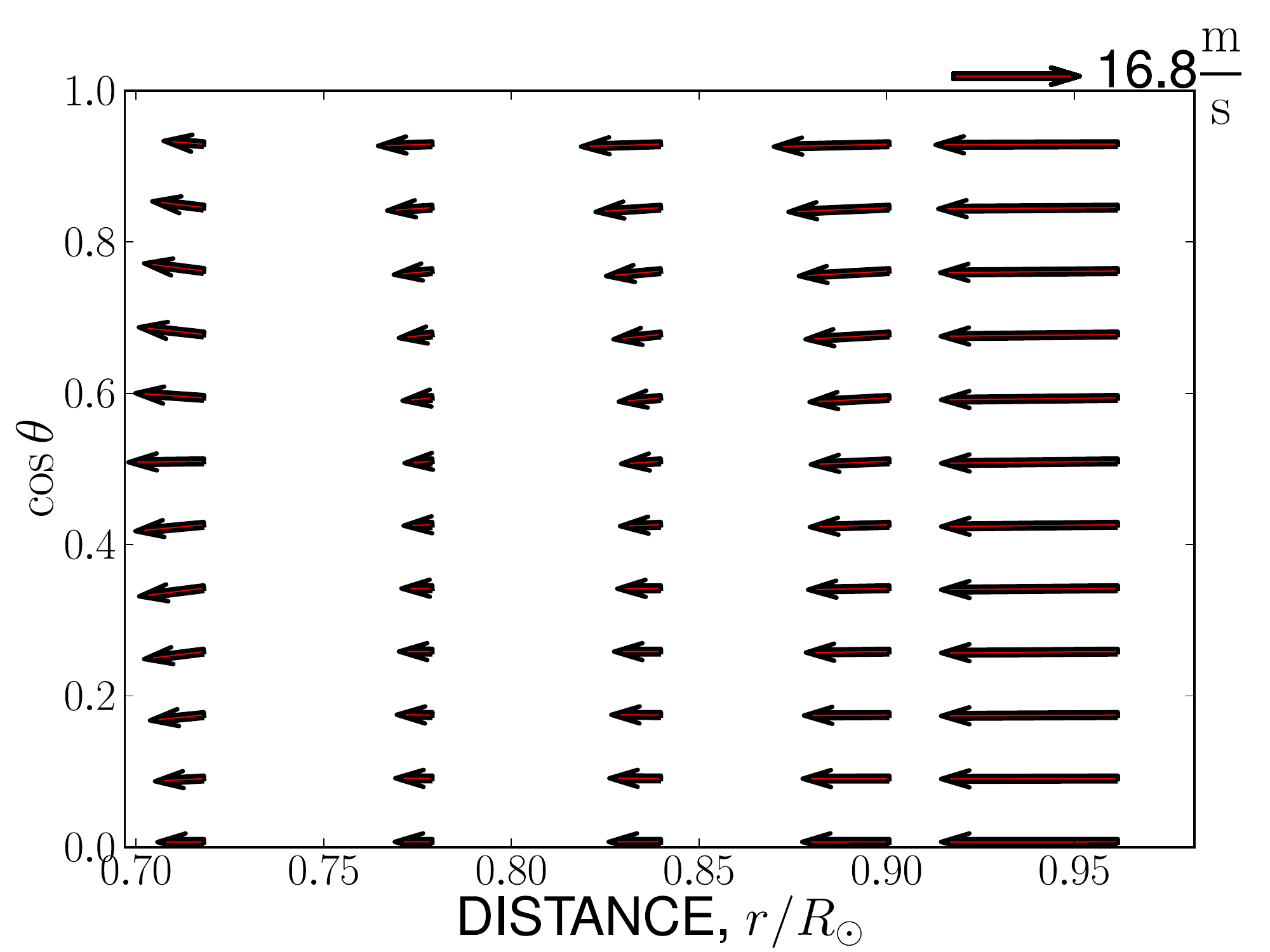}
\par\end{centering}

\caption{\label{fig:sum-pump}The patterns of the total (including the diamagnetic
and the density gradient effects) pumping velocity fields for the
toroidal magnetic field(left) and for the poloidal field(right).}
\end{figure}
The expressions given by Eq.~(\ref{eq:EMFP}) are valid for the case of
weak shear, when $\tau_{c}\max\left(\left|\nabla_{i}\overline{V}_{j}\right|\right)\ll1$.
In terms of the strain tensor $\bm{\tilde{V}}$
this condition of weak shear implies ${\displaystyle \varOmega^{\star}}\max\left(\left|r\partial_{r}\tilde{\varOmega}\right|,\left|\partial_{\theta}\tilde{\varOmega}\right|\right)\ll1$.
This is not valid at the bottom of the solar convection zone
where the radial gradient of the angular velocity is strong and $\varOmega^{\star}\gg1$
and $\tau_{c}\max\left(\left|\nabla_{i}\overline{V}_{j}\right|\right)\approx2$.
\citet{2010GApFD.104..167L} suggested that this pumping effect
is quenched with increasing shear inversely proportional to $\left(\tau_{c}\max\left(\left|\nabla_{i}\overline{V}_{j}\right|\right)\right)^{1\dots2}$.
Therefore, we introduce an ad-hoc quenching function for the pumping
effect:
\begin{equation}
f^{(S)}=\frac{1}{1+C_{S}{\displaystyle \varOmega^{*s}\left(\left|r\frac{\partial\tilde{\varOmega}}{\partial r}\right|+\left|\frac{\partial\tilde{\varOmega}}{\partial\theta}\right|\right)^{s}}},
\end{equation}
where $C_{S}$ is a constant to control the magnitude of the quenching,
and $s=1$. Results by \citet{2010GApFD.104..167L} suggest $1 < s < 2$
in relation to geometry of the large-scale shear.
We find that for the solar convection zone the amplitude
of the pumping effect does not change very much ($\sim1\,\mathrm{m/s}$)
with $s$ varying in the range $1\dots2$.

From the given relations, using $\EMF^{(H)}=\bm{U}^{\rm(eff)}\times\bm{\overline{B}}$,
we find the effective drift velocity, $\bm{U}^{\rm(eff)}$, due to
the helicity--vorticity pumping effect. 
Taking into account the variation of turbulence parameters in
the solar convection zone we compute $\bm{U}^{\rm(eff)}$. The bottom panel
of Figure \ref{fig:CZ} shows the distribution of the velocity field
$\bm{U^{(eff)}}$ for the helicity--vorticity pumping effect for
the toroidal and poloidal components of the large-scale magnetic field.
The maximum velocity drift occurs in the middle and at
the bottom of the convection zone. The direction of drift has equatorial
and polar cells corresponding to two regions in the solar convection
zone with different signs of the radial gradient of the angular velocity.
The anisotropy in transport of the toroidal and poloidal components
of the large-scale magnetic field is clearly seen. 

The other important pumping effects are due to mean density and
turbulence intensity gradients \citep{zeld57,kit:1991,kit-rud:1993a,2001ApJ...549.1183T}.
These effects were estimated using Eq.~(\ref{eq:pump}). For these
calculations we put $C_{\eta}=1$, $\varepsilon=0$, $C_{\mathcal{K}}=1$
and $\overline{\chi}=0$. Figure \ref{fig:sum-pump} shows the sum
of the pumping effects for the toroidal and poloidal components of
mean magnetic fields including the helicity--vorticity pumping
effect. In agreement with previous studies, it is found that the
radial direction is the principal direction of mean-field transport
in the solar convection zone. In its upper part the transport
is downward because of pumping due to the density gradient \citep{kit:1991}.
At the bottom of the convection zone the diamagnetic pumping effect
produces downward transport as well \citep{kit:1991,1995A&A...296..557R}.
The diamagnetic pumping is quenched inversely proportional to the Coriolis number
(e.g., \citealp{kit:1991,pi08Gafd}) and it has the same order
of magnitude as the helicity--vorticity pumping
effect. The latter effect modifies the direction of effective drift of the
toroidal magnetic field near the bottom of the convection zone. There
is also
upward drift of the toroidal field at low latitudes in the middle
of the convection zone. It results from the combined effects of
density gradient and global rotation \citep{kit:1991,2004IAUS..223..277K}.
For the poloidal magnetic field the transport is downward everywhere
in the convection zone. At the bottom of the convection zone the action
of the diamagnetic pumping on the meridional component of the large-scale
magnetic field is amplified due to the helicity--vorticity pumping
effect. 

The obtained pattern of large-scale magnetic field drift
in the solar convection zone does not take into account nonlinear
effects, e.g., because of magnetic buoyancy. The effect of
mean-field buoyancy is rather small compared with flux-tube buoyancy
(\citealp{1993A&A...274..647K}, cf.\ \citealp{2011A&A...533A..40G}).

To find out the current helicity counterpart of the pumping effect
we analyze dynamo models by solving Eqs.~(\ref{eq:dyn}, \ref{eq:hel}).
The governing parameters of the model are $C_{\eta}=0.05$, $C_{\omega}={\displaystyle \frac{1}{3}C_{\alpha}}$.
We discuss the choice of the governing parameters later. The other
parameters of the model are given in the Table \ref{tab:The-parameters-of}.
Because of the weakening factor $C_{\eta}$ the magnitude of the pumping
velocity is about one order of magnitude smaller than what is shown in Figure~\ref{fig:sum-pump}. 

Following \citet{pk11apjl}, we use a combination of ``open''
and ``closed'' boundary conditions at the top, controlled by a parameter
$\delta=0.95$, with
\begin{equation}
\delta\frac{\eta_{T}}{r_{e}}B+\left(1-\delta\right)\mathcal{E}_{\theta}=0.\label{eq:tor-vac}
\end{equation}
This is similar to the boundary condition discussed by \citet{kit:00}.
For the poloidal field we apply a combination of the local condition
$A=0$ and the requirement of a smooth transition from the internal poloidal
field to the external potential (vacuum) field: 
\begin{equation}
\delta\left(\left.\frac{\partial A}{\partial r}\right|_{r=r_{e}}-\left.\frac{\partial A^{(vac)}}{\partial r}\right|_{r=r_{e}}\right)+\left(1-\delta\right)A=0.\label{eq:pol-vac}
\end{equation}
We assume perfect conductivity at the bottom boundary with standard
boundary conditions. For the magnetic helicity, similar to \citet{guero10},
we put $\boldsymbol{\nabla}_{r}\bar{\chi}=0$ at the bottom of the
domain and $\bar{\chi}=0$ at the top of the convection zone.

\begin{table}

\caption{\label{tab:The-parameters-of}The parameters of the models. Here,
$B_{\max}$ is the maximum of the toroidal magnetic field strength inside
the convection zone, $P$ is the dynamo period of the model.}

\begin{tabular}{|c|c|c|c|c|c|c|}
\hline 
Model & $C_{\alpha}$ & $R_{\chi}$ & ${\displaystyle C_{\mathcal{K}}}$ & ${\displaystyle C_{\mathcal{C}}}$ & $B_{\max}$ [G] & $P$ [yr] \tabularnewline
\hline 
\hline 
D1 & 0.025 & $10^{3}$ & 0 & 0 & 500 & 16\tabularnewline
\hline 
D2 & 0.025 & $10^{3}$ & 1 & 0 & 250 & 13\tabularnewline
\hline 
D3 & 0.03 & $10^{3}$ & 1 & 10 & 300 & 13\tabularnewline
\hline 
D4 & 0.035 & $5\cdot10^{2}$ & 1 & 1 & 500 & 11\tabularnewline
\hline 
\end{tabular}

\end{table}

In this paper we study dynamo models which include R\"adler's
$\boldsymbol{\varOmega}\times\bm{J}$ dynamo effect due to a
large-scale current  and global rotation \citep{rad69}. There is also
a dynamo effect due to large-scale shear and current
\citep{kle-rog:04a}.
 The motivation to consider these addional turbulent sources in the
 mean-field dynamo comes
from DNS dynamo experiments \citep{2007NJPh....9..305B,2008A&A...491..353K,2009PhRvL.102d4501H,2009A&A...500..633K}
and from our earlier studies \citep{2009A&A...493..819P,2009A&A...508....9S}.
The dynamo effect due to large-scale current gives an additional
source of large-scale poloidal magnetic field. This can
help to solve the issue with the dynamo period being otherwise too short. Also, in the models the
large-scale current dynamo effect produces less overlapping cycles
than dynamo models with $\alpha$-effect alone. The choice
of parameters in the dynamo is justified by our previous studies
\citep{2009A&A...493..819P,pk11mf}, where we showed that solar-types
dynamos can be obtained for $C_{\alpha}/C_{\omega}>2$. In those
papers we find the approximate threshold to be $C_{\alpha}\approx0.02$
for a given diffusivity dilution factor of $C_{\eta}=0.05$. 

\begin{figure}
a)\includegraphics[width=0.85\textwidth,height=3.5cm]{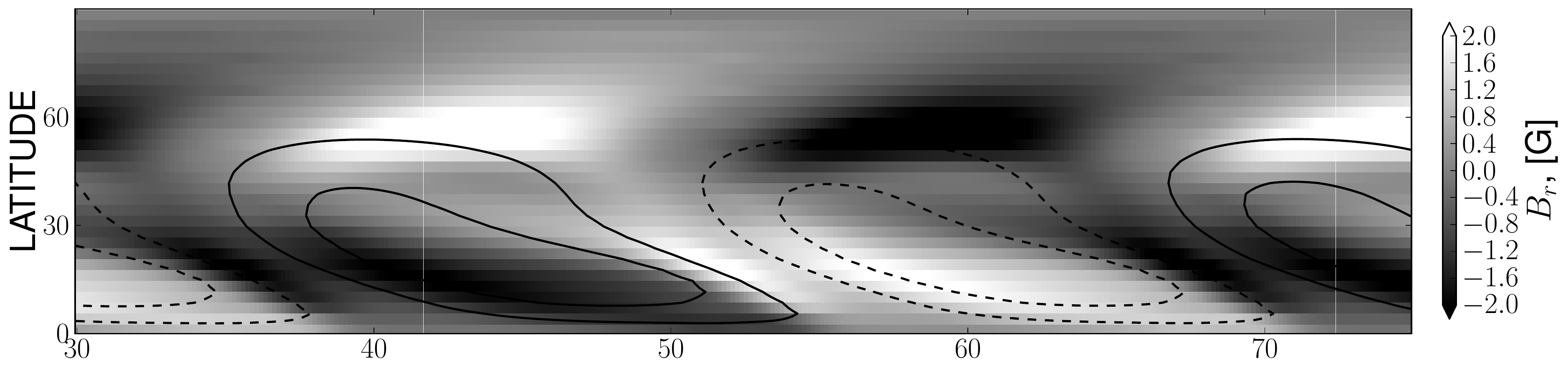}

b)\includegraphics[width=0.8\textwidth,height=3.5cm]{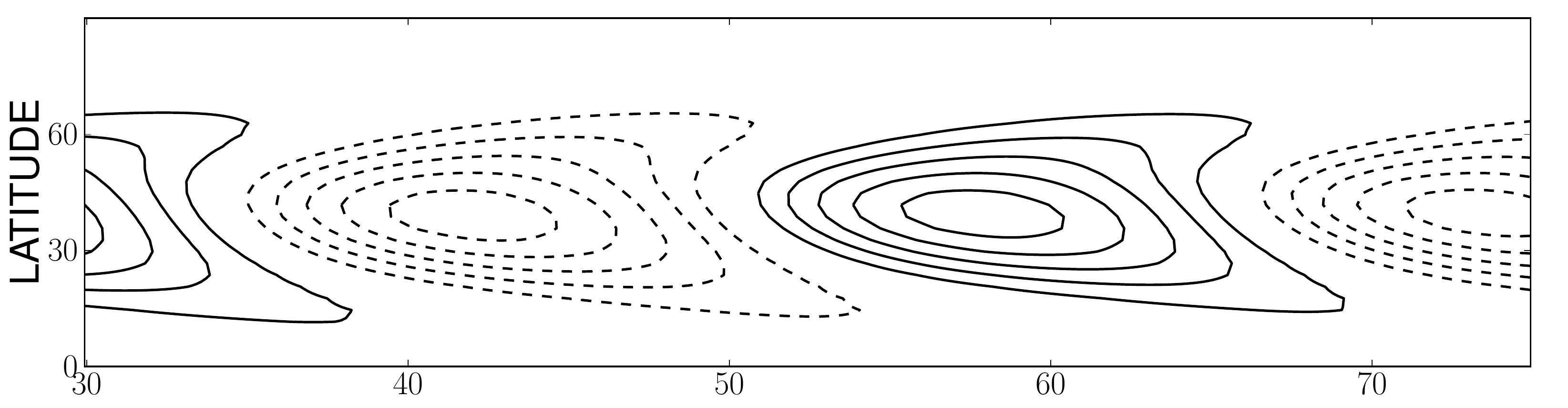}

c)\includegraphics[width=0.85\textwidth,height=3.5cm]{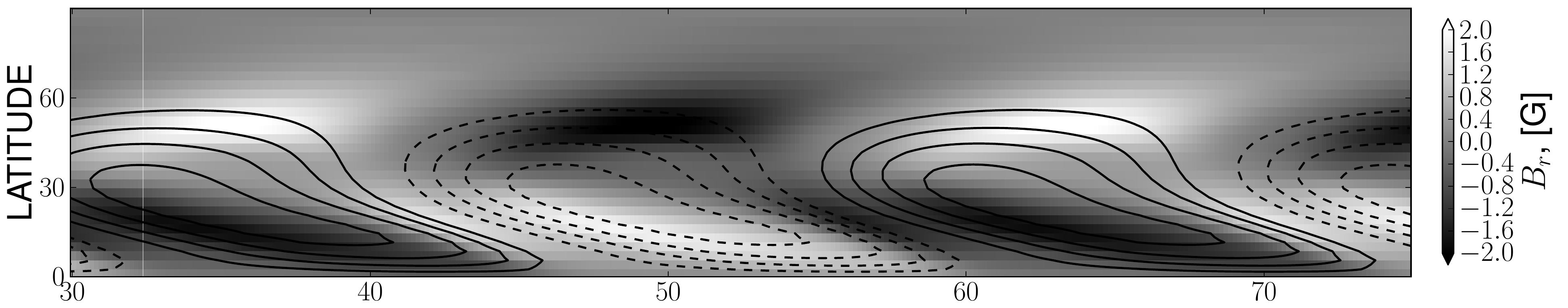}

d)\includegraphics[width=0.8\textwidth,height=3.5cm]{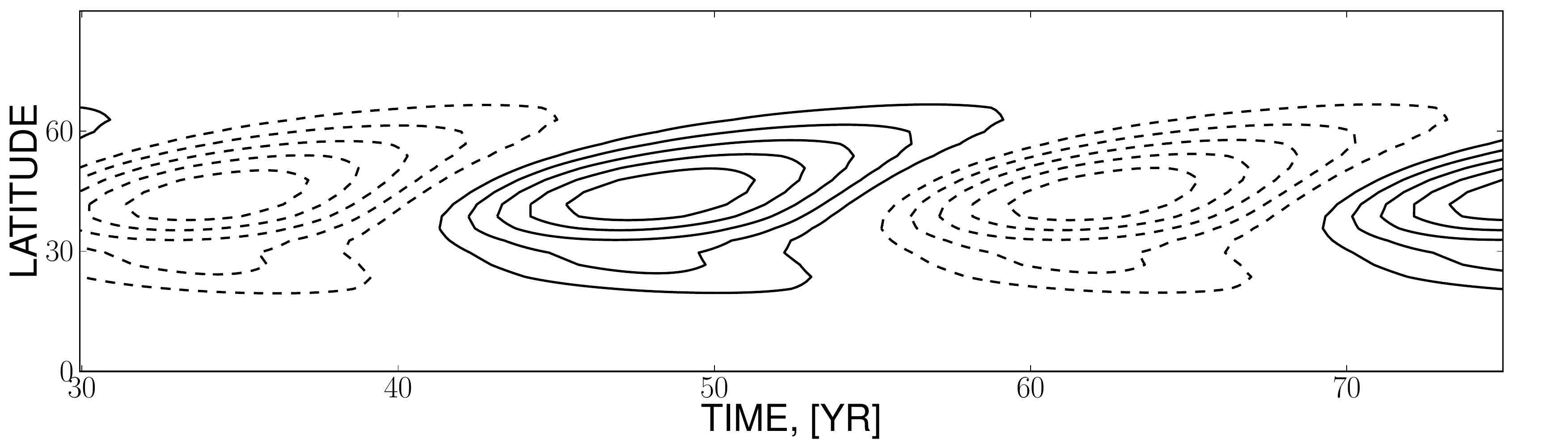}

\caption{\label{evo1}The time-latitude diagrams for the toroidal and radial
magnetic fields for the models D1 and D2: a) the model D1, the toroidal
field (iso-contours, $\pm.25KG$) near the surface and the radial field
(gray-scale density plot); b) the model D1, the toroidal field at
the bottom of the solar convection zone, the contours drawn in the
range $\pm.5KG$; c) the same as for item a) for the model D2; d)
the same as for item b) for the model D2.}
\end{figure}

As follows from the results given in Fig.\ref{fig:sum-pump}, the kinetic
helicity--vorticity pumping effect has a negligible contribution in the
near-surface layers, where downward pumping due to density stratification
dominates. Therefore, it is expected that the surface dynamo waves
are not affected if we discard magnetic helicity from the dynamo
equations. Figure \ref{evo1} shows time-latitude diagrams for
toroidal and radial magnetic fields at the surface and for toroidal
magnetic field at the bottom of the convection zone for two dynamo
models D1 and D2 with and without the helicity--vorticity pumping
effect, but magnetic helicity is taken into account as the main dynamo
quenching effect. To compare with observational data from a time-latitude
diagram of sunspot area (e.g., \citealp{2011SoPh..273..221H}),
we multiply the toroidal field component $B$ by factor $\sin\theta$.
This gives a quantity, which is proportional to the flux of large-scale
toroidal field at colatitude $\theta$. We further assume that
the sunspot area is related to this flux.

Near the surface, models D1 and D2 give similar patterns
of magnetic field evolution. At the bottom of the convection zone
model D1 shows both poleward and equatorward branches of the
dynamo wave propagation that is in agreement with the Parker-Yoshimura
rule. Both branches have nearly the same time scale that equals $\simeq$16
years. The results from model D2 show that at the bottom of the
convection zone the poleward branch of the dynamo wave dominates.
Thus we conclude that the helicity--vorticity pumping effect alters the
propagation of the dynamo wave near the bottom of the solar convection
zone. We find that models with magnetic helicity contributions
to the pumping effect do not change this conclusion.

\begin{figure}
\includegraphics[width=0.5\textwidth]{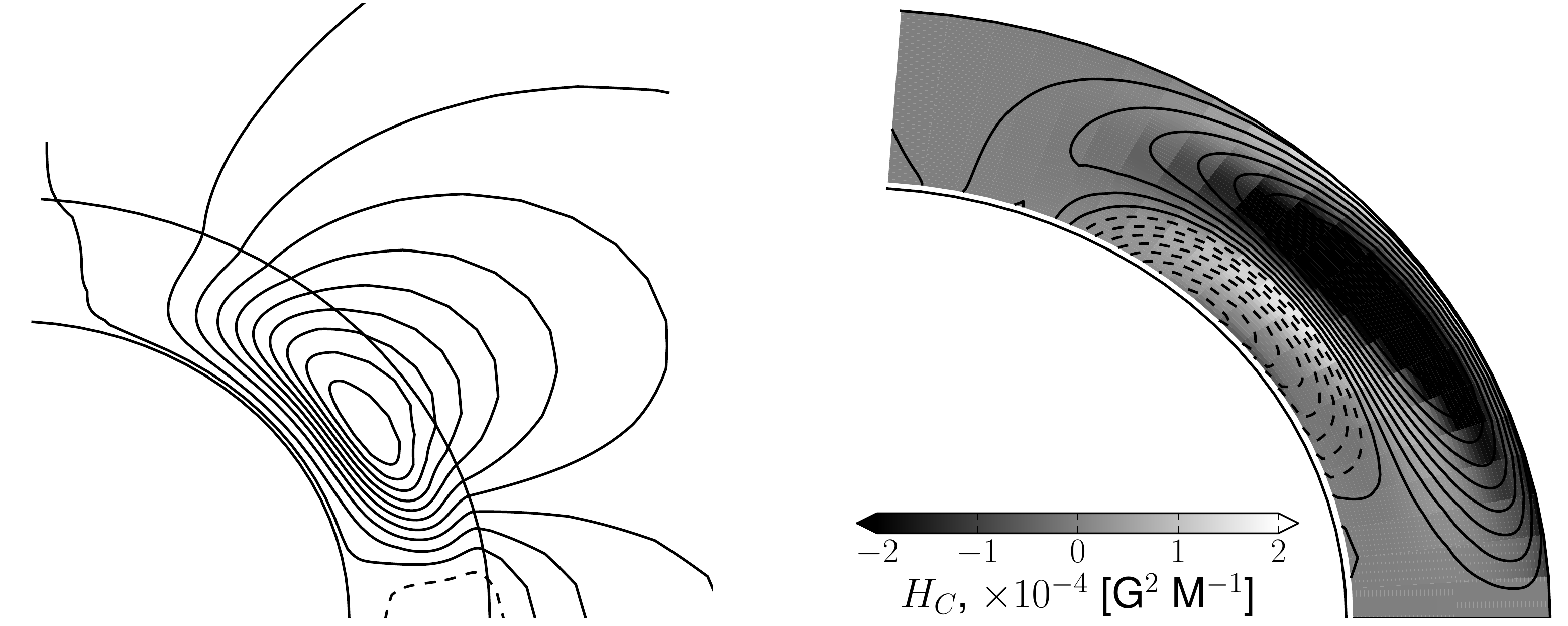}

\caption{\label{fig:Snapshot-for-the}Snapshots for the mean magnetic field
and the current helicity distributions at the north hemisphere in
the model D4. Left panel shows the field lines of the poloidal
component of the mean magnetic field. The right panel shows the toroidal
magnetic field (iso-contours $\pm500$G) and the current helicity
(gray scale density plot).}
\end{figure}

Figure \ref{fig:Snapshot-for-the} shows a typical snapshot of
the magnetic helicity distribution in the northern hemisphere for all
our models. The helicity has a negative sign in the bulk of the
solar convection zone. Regions with positive current
helicity roughly correspond to domains of the negative large-scale current helicity concentration.
They are located in the middle of the solar convection zone and at
the high and low latitudes near the top of the solar convection zone.
As follows from Fig.~\ref{fig:Snapshot-for-the}, the pumping effect
due to current helicity may be efficient in the upper part of
the solar convection zone where it might intensify the equatorial
drift of the dynamo wave along iso-surfaces of the angular velocity.
We find that the pumping effect that results from magnetic
helicity is rather small in our models. This may be due to the weakness of the
magnetic field.
Observations \citep{zetal10}
give about one order magnitude larger current helicity than what
is shown in Fig.~\ref{fig:Snapshot-for-the}. In the model we
estimate the current helicity as $H_{\mathcal{C}}={\displaystyle \frac{\overline{\chi}}{\mu_{0}\ell^{2}}}$.
This result depends essentially on the mixing length parameter $\ell$.
The stronger helicity is concentrated to the surface, the larger $H_{\mathcal{C}}$.
In observations, we do not know from were the helical magnetic structures
come from. In view of the given uncertainties we estimate the probable
effect of a larger magnitude of magnetic helicity in the model
by increasing the parameter $C_{\mathcal{C}}$ to 10 (model D3).
In addition, we consider the results for the nonlinear model D4. It has
a higher $C_{\alpha}$ and a lower $R_{\chi}$ to increase the nonlinear
impact of the magnetic helicity on the large-scale magnetic field
evolution. 

\begin{figure}
\includegraphics[width=0.85\textwidth,height=3.5cm]{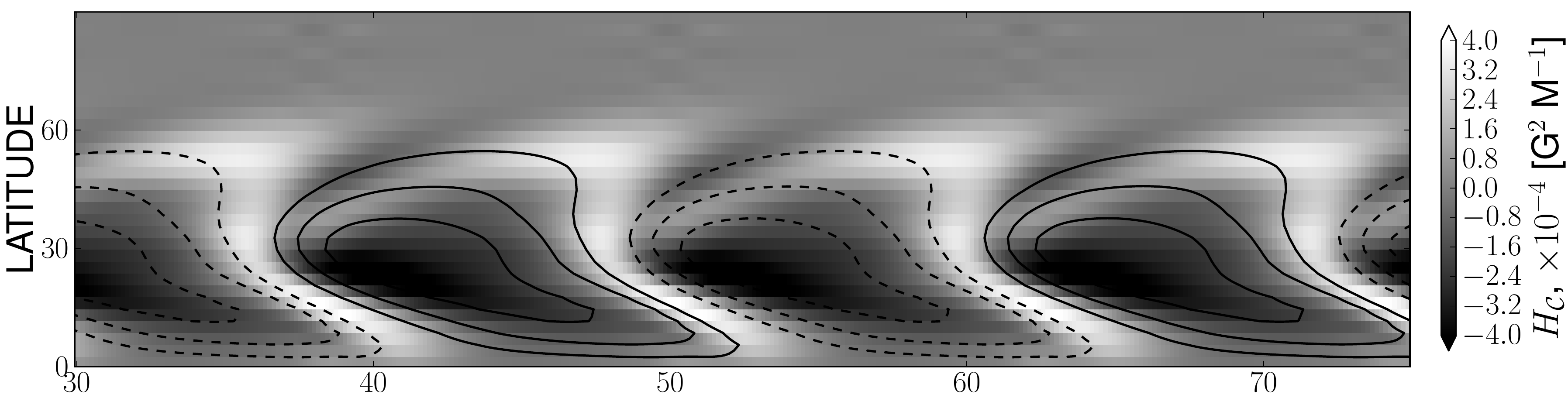}

\includegraphics[width=0.85\textwidth,height=3.5cm]{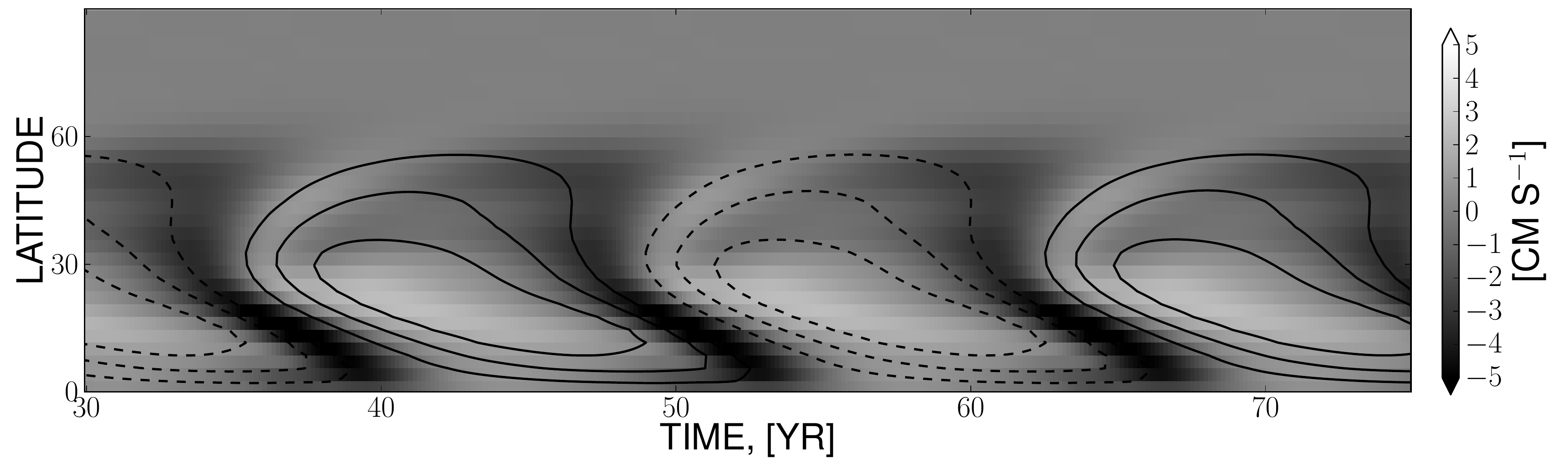}

\caption{\label{evo2} Top, the near-surface time-latitude diagrams for the
toroidal magnetic field and the current helicity for the models D4.
Bottom, the near-surface time-latitude diagrams for the toroidal magnetic
field and the latitudinal component of the drift velocity induced by the magnetic
helicity for the model D3.}
\end{figure}

The top panel of Figure~\ref{evo2} shows a time-latitude diagram
of toroidal magnetic field and current helicity evolution
near the surface for model D4. We find a positive sign of
current helicity at the decay edges of the toroidal magnetic field
butterfly diagram. There are also areas with positive
magnetic helicity at high latitudes at the growing edges of the
toroidal magnetic field butterfly diagram. The induced pumping velocity
is about 1 $\mathrm{cm\, s^{-1}}$. The increase of the magnetic helicity
pumping effect by a factor of 10 (model D3) shifts the latitude of the
maximum of the toroidal magnetic field by about $5^{\circ}$ toward the equator.
The induced pumping velocity is about 5 $\mathrm{cm\, s^{-1}}$. 

Stronger nonlinearity (model D4) and a stronger magnetic
helicity pumping effect (model D3) modify the butterfly diagram in
different ways. Model D3 shows a simple shift of the maximum
of toroidal magnetic field toward the equator. Model D4 shows a
fast drift of large-scale toroidal field at the beginning of a
cycle and a slow-down of the drift velocity as the cycle progresses.

\begin{figure}
\includegraphics[width=0.5\textwidth]{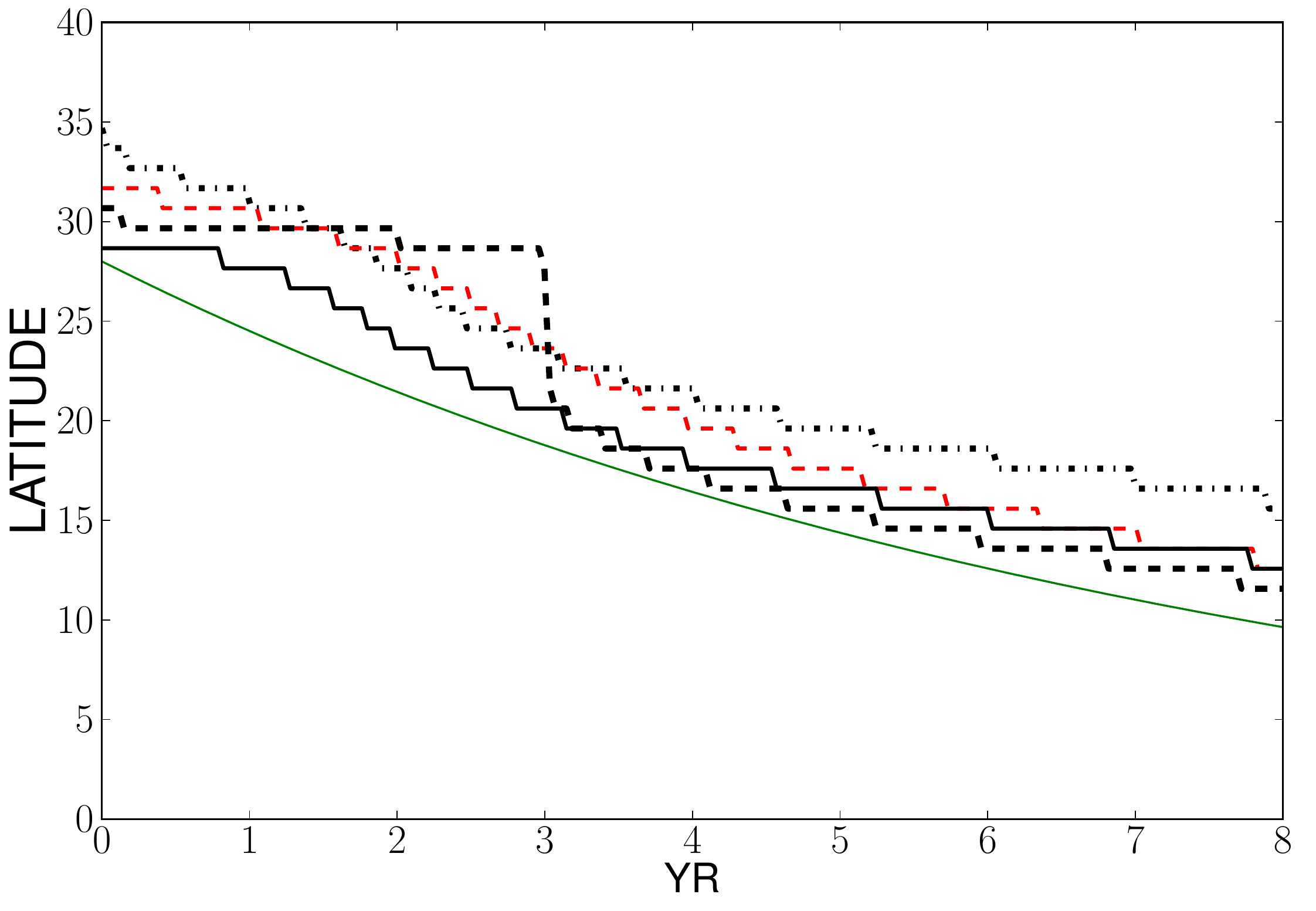}\includegraphics[width=0.5\textwidth]{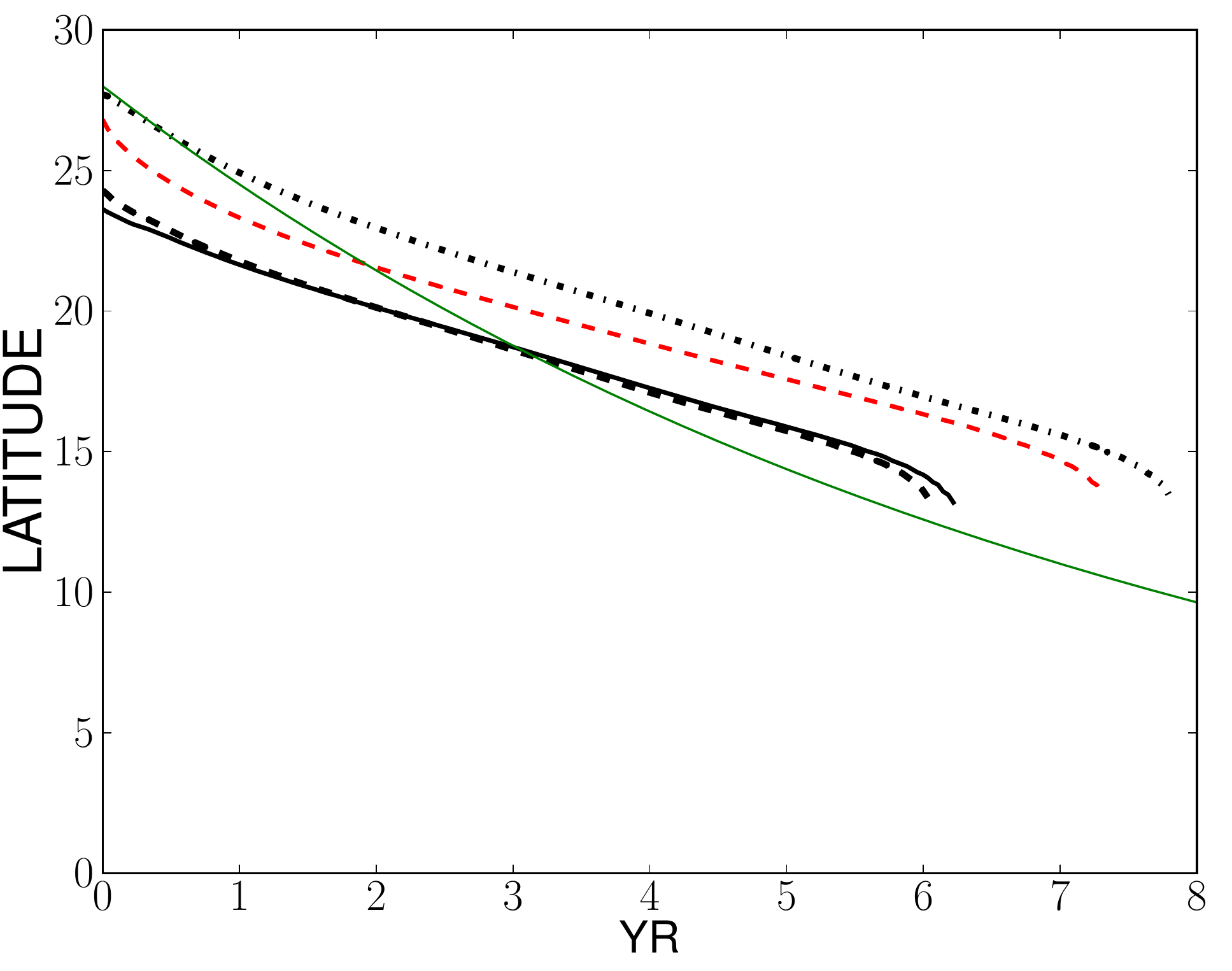}

\caption{\label{fig:The-latitude-of}The drift of the latitude of maximum (left)
and the centroid position of the magnetic flux at the near-surface
layer in the models D(1-4). The dash-dotted line shows results for
the model D1, the red dashed line - for the model D2, the solid black
line - for the model D3, the black dashed line - for the model D4
and the solid green line shows the exponential law of the sunspot area
centroid drift, as suggested by \citet{2011SoPh..273..221H}. }

\end{figure}

Figure \ref{fig:The-latitude-of} shows in more detail the latitudinal
drift of the maximum of the toroidal magnetic field evolution during
the cycle (left panel in the Figure \ref{fig:The-latitude-of}), 
\begin{equation}
\lambda_{\max}(t)=90^{\circ}-\underset{\theta>45^{\circ}}{\max}\left(\left|B_{S}(\theta)\right|\sin\theta\right),\label{eq:lmax}
\end{equation}
and the latitudinal drift of the centroid position of the toroidal
magnetic field flux (cf.\ \citealp{2011SoPh..273..221H}) 
\begin{equation}
\lambda_{C}(t)=90^{\circ}-\frac{\int_{0}^{\pi/2}\theta B_{S}(\theta)\sin\theta d\theta}{\int_{0}^{\pi/2}B_{S}(\theta)\sin\theta d\theta},\label{eq:lcen}
\end{equation}
where $B_{S}(\theta)={\displaystyle \left\langle B\left(r,\theta\right)\right\rangle _{(0.9,0.99)R}}$
is the toroidal magnetic field, which is averaged over the surface
layers. Note that the overlap between subsequent cycles influences
the value of $\lambda_{C}$ more than the value of $\lambda_{\max}$. The
behaviour of $\lambda_{\max}$ in models D1,D2 and D3 reproduces
qualitatively the exponential drift of maximum latitude as suggested
by \citet{2011SoPh..273..221H}:
\[
\lambda_{C}(t)=28^{\circ}\exp\left(-\frac{12t}{90}\right),
\]
where $t$ is time measured in years. Model D4 shows a change
between fast (nearly steady dynamo wave) drift at the beginning
of the cycle to slow drift at the decaying phase of the cycle.
The overlap between subsequent cycles is growing from model
D1 to model D4. 
 In all the models the highest latitude of the centroid position of
the toroidal magnetic flux is below 30$^{\circ}$. Models
D3 and D4 have nearly equal starting latitude of the centroid position.
It is about 24$^{\circ}$. This means that a model with increased
magnetic helicity pumping produces nearly the same effect for the
shift of the centroid position as a model with a strong nonlinear
effect of magnetic helicity.

\section{Discussion and conclusions}

We have shown that the interaction of helical convective motions and
differential rotation in the solar convection zone produces a turbulent
drift of large-scale magnetic field. The principal direction of
the drift corresponds to the direction of the large-scale vorticity vector.
The large-scale vorticity vector
roughly follows to iso-surfaces of angular velocity. Since the
direction of the drift depends on the sign of helicity, the
pumping effect is governed by the Parker-Yoshimura rule \citep{par55,yosh1975}. 

The effect is computed within the framework of mean-field magnetohydrodynamics
using the minimal $\tau$-approximation. In the calculations, we have assumed
that the turbulent kinetic and current helicities are given. The calculations
were done for arbitrary Coriolis number. In agreement with \citet{mitr2009AA}
and \citet{garr2011}, the analytical calculations show that the leading
effect of pumping is described by a large-scale magnetic
drift in the direction of the large-scale vorticity vector and by
anisotropic pumping which produces a drift of toroidal and poloidal
components of the field in opposite directions. The component of the
drift that is induced by global rotation and helicity (second
line in Eq.~(\ref{eq:EMFP})) is rather small compared to the main
effect. The latter conclusion should be checked separately for a
different model of background turbulence, taking into account
the generation of kinetic helicity due to global rotation and
stratification in a turbulent medium. 

We have estimated the pumping effect for the solar convection zone and
compared it with other turbulent pumping effects including
diamagnetic pumping and turbulent pumping that results from magnetic
fluctuations in stratified turbulence \citep{kit:1991,pi08Gafd}.
The latter is sometimes referred to as ``density-gradient pumping effect''
\citep{2004IAUS..223..277K}. The diamagnetic pumping is upward in
the upper part of the convection zone and downward near the bottom.
The velocity field of density-gradient pumping is more complicated
(see Figure 4). However, its major effect is concentrated near the
surface. Both diamagnetic pumping and density-gradient pumping
effects are quenched inversely proportional to the Coriolis number \citep{kit:1991,pi08Gafd}.
The helicity--vorticity pumping effect modifies the direction of
large-scale magnetic drift at the bottom of the convection zone.
This effect was illustrated by a dynamo model that shows a dominant
poleward branch of the dynamo wave at the bottom of the convection
zone.

It is found that the magnetic helicity contribution of the pumping
effect can be important for explaining the fine structure of the
sunspot butterfly diagram. In particular, the magnetic helicity contribution
results in a slow-down of equatorial propagation of the dynamo wave.
The slow-down starts just before the maximum of the cycle. Observations
indicate a similar behavior in sunspot activity \citep{ternul2007AN,2011SoPh..273..221H}.
A behavior like this can be seen in flux-transport models as
well (\citealp{2006ApJ...647..662R}). For the time being it is unclear
what are the differences between different dynamo
models and how well do they reproduce the observations. A more detailed
analysis is needed.

\bibliographystyle{plainnat}

\section*{Appendix A}

To compute $\EMF$ it is convenient to write equations (\ref{induc1})
and (\ref{navie1}) in Fourier space: 
\begin{eqnarray}
\left(\frac{\partial}{\partial t}+\eta z^{2}\right)\hat{b}_{j} & = & \frac{i}{\mu}z_{l}\hat{u}_{j}\left(\bm{z}\right)\bar{B}_{l}+\label{induc2}\\
 & + & iz_{l}\int\left[\hat{b}_{l}(\bm{z-q)}\widehat{\bar{V}}_{j}(\bm{q})-\hat{b}_{j}\left(\bm{z-q)}\widehat{\bar{V}}_{l}(\bm{q})\right]d\bm{q}+\widehat{\mathfrak{G}}_{j}.\right.\nonumber \\
\left(\frac{\partial}{\partial t}+\nu z^{2}\right)\hat{u}_{i} & = & \hat{f}_{i}+\widehat{\mathfrak{F}}_{i}-2\left(\bm{\varOmega}\widehat{\bm{z}}\right)\left(\widehat{\bm{z}}\times\widehat{\bm{u}}\right)_{i}\label{navie2}\\
 & - & i\pi_{if}(\bm{z)}z_{l}\int\left[\hat{u}_{l}(\bm{z-q)}\widehat{\bar{V}}_{f}(\bm{q})+\hat{u}_{f}\left(\bm{z-q)}\widehat{\bar{V}}_{l}(\bm{q})\right]{d}\bm{q}\right.\nonumber \\
 & + & \frac{i}{\mu}\hat{b}_{i}\left(\bm{z}\right)\left(\bm{z}\ccdot\bar{\bm{B}}\right),\nonumber 
\end{eqnarray}
where the turbulent pressure was excluded from (\ref{navie1}) by
convolution with tensor $\pi_{ij}(\bm{z)}=\delta_{ij}-\hat{z}_{i}\hat{z}_{j}$,
$\delta_{ij}$ is the Kronecker symbol and $\widehat{\bm{z}}$
is a unit wave vector. The equations for the second-order moments
that make contributions to the mean-electro-motive force(MEMF) can be found directly from (\ref{induc2},
\ref{navie2}). As the preliminary step we write the equations for
the second-order products of the fluctuating fields, and make the
ensemble averaging of them, 
\begin{eqnarray}
\frac{\partial}{\partial t}\left\langle \hat{u}_{i}\left(\bm{z}\right)\hat{b}_{j}\left(\bm{z}'\right)\right\rangle  & = & i\, z'_{l}\bar{B}_{l}\left\langle \hat{u}_{i}\left(\bm{z}\right)\hat{u}_{j}\left(\bm{\left.z'\right)}\right\rangle \right.-2\left(\bm{\varOmega}\widehat{\bm{z}}\right)\varepsilon_{iln}\widehat{z_{l}}\left\langle \hat{u}_{n}(\bm{z})\hat{b}_{j}(\bm{z'})\right\rangle \label{eq:kappa1}\\
 & + & i\, z'_{l}\int\left[\left\langle \hat{u}_{i}\left(\bm{z}\right)\hat{b}_{l}\left(\bm{z'-q)}\right\rangle \widehat{\bar{V}}_{j}(\bm{q})-\left\langle \hat{u}_{i}\left(\bm{z}\right)\hat{b}_{j}\left(\bm{z'-q)}\right\rangle \widehat{\bar{V}}_{l}(\bm{q})\right]d\bm{q}\right.\right.\nonumber \\
 & - & i\,\pi_{if}(\bm{z)}z_{l}\int\left[\left\langle \hat{u}_{l}\left(\bm{z-q)}\hat{b}_{j}\left(\bm{z}'\right)\right\rangle \widehat{\bar{V}}_{f}(\bm{q})+\left\langle \hat{u}_{f}\left(\bm{z-q)}\hat{b}_{j}\left(\bm{z}'\right)\right\rangle \widehat{\bar{V}}_{l}(\bm{q})\right]d\bm{q}\right.\right.\nonumber \\
 & + & \frac{i}{\mu}\bar{B}_{l}z_{l}\langle\hat{b}_{i}\left(\bm{\left.z\right)}\hat{b}_{j}\left(\bm{z}'\right)\rangle+Th_{ij}^{\varkappa}(\bm{z},\bm{z'})-\left(\eta z'^{2}+\nu z^{2}\right)\left\langle \hat{u}_{i}\left(\bm{z}\right)\hat{b}_{j}\left(\bm{z}'\right)\right\rangle ,\right.\nonumber \\
\frac{\partial}{\partial t}\left\langle \hat{u}_{i}\left(\bm{z}\right)\hat{u}_{j}\left(\bm{z}'\right)\right\rangle  & = & -2\left(\bm{\varOmega}\widehat{\bm{z}}\right)\varepsilon_{iln}\widehat{z_{l}}\left\langle \hat{u}_{n}(\bm{z})\hat{u}_{j}(\bm{z'})\right\rangle -2\left(\bm{\varOmega}\widehat{\bm{z}}'\right)\varepsilon_{jln}\widehat{z_{l}'}\left\langle \hat{u}_{i}(\bm{z})\hat{u}_{n}(\bm{z'})\right\rangle \label{secm1}\\
 & - & i\,\pi_{if}(\bm{z)}z_{l}\int\left[\left\langle \hat{u}_{l}\left(\bm{z-q)}\hat{u}_{j}\left(\bm{z}'\right)\right\rangle \widehat{\bar{V}}_{f}(\bm{q})+\left\langle \hat{u}_{f}\left(\bm{z-q)}\hat{u}_{j}\left(\bm{z}'\right)\right\rangle \widehat{\bar{V}}_{l}(\bm{q})\right]d\bm{q}\right.\right.\nonumber \\
 & - & i\,\pi_{jf}(\bm{z')}z'_{l}\int\left[\left\langle \hat{u}_{i}\left(\bm{z}\right)\hat{u}_{l}\left(\bm{z-q)}\right\rangle \widehat{\bar{V}}_{f}(\bm{q})+\left\langle \hat{u}_{i}\left(\bm{z}\right)\hat{u}_{f}\left(\bm{z-q)}\right\rangle \widehat{\bar{V}}_{l}(\bm{q})\right]d\bm{q}\right.\right.\nonumber \\
 & + & Th_{ij}^{v}(\bm{z},\bm{z'})-\nu\left(z'^{2}+z^{2}\right)\left\langle \hat{u}_{i}\left(\bm{z}\right)\hat{u}_{j}\left(\bm{z}'\right)\right\rangle ,\nonumber \\
\frac{\partial}{\partial t}\left\langle \hat{b}_{i}\left(\bm{z}\right)\hat{b}_{j}\left(\bm{z}'\right)\right\rangle  & = & Th_{ij}^{h}(\bm{z},\bm{z'})-\left(\eta z'^{2}+\eta z^{2}\right)\left\langle \hat{b}_{i}\left(\bm{z}\right)\hat{b}_{j}\left(\bm{z}'\right)\right\rangle \label{eq:mag1}\\
 & + & i\, z'_{l}\int\left[\left\langle \hat{b}_{i}\left(\bm{z}\right)\hat{b}_{l}\left(\bm{z'-q)}\right\rangle \widehat{\bar{V}}_{j}(\bm{q})-\left\langle \hat{b}_{i}\left(\bm{z}\right)\hat{b}_{j}\left(\bm{z'-q)}\right\rangle \widehat{\bar{V}}_{l}(\bm{q})\right]d\bm{q},\right.\right.\nonumber 
\end{eqnarray}
 where, the terms $Th_{ij}^{(\varkappa,v,h)}$ involve the third-order
moments of fluctuating fields and second-order moments of them with
the forcing term. Next, we apply the $\tau$-approximation, substituting
the $Th_{ij}^{(\varkappa,v,h)}$-terms by the corresponding $\tau$
relaxation terms of the second-order contributions,

\begin{align}
Th_{ij}^{(\varkappa)} & \rightarrow-\left\langle \hat{m}_{i}\left(\bm{z}\right)\hat{b}_{j}\left(\bm{z}'\right)\right\rangle /\tau_{c},\label{eq:thx-1}\\
Th_{ij}^{(v)} & \rightarrow-\frac{\left\langle \hat{m}_{i}\left(\bm{z}\right)\hat{m}_{j}\left(\bm{z}'\right)\right\rangle -\left\langle \hat{m}_{i}\left(\bm{z}\right)\hat{m}_{j}\left(\bm{z}'\right)\right\rangle ^{\left(0\right)}}{\tau_{c}},\label{eq:thv-1}\\
Th_{ij}^{(h)} & \rightarrow-\frac{\left\langle \hat{b}_{i}\left(\bm{z}\right)\hat{b}_{j}\left(\bm{z}'\right)\right\rangle -\left\langle \hat{b}_{i}\left(\bm{z}\right)\hat{b}_{j}\left(\bm{z}'\right)\right\rangle ^{(0)}}{\tau_{c}},\label{eq:thh-1}
\end{align}
where the superscript $\dots^{\left(0\right)}$ denotes the moments
of the background turbulence. Approximating these complicated contributions
by the simple relaxation terms has to be considered as a questionable
assumption. It involves additional assumptions (see \citealp{2007GApFD.101..117R}),
e.g., it is assumed that the second-order correlations in Eq.~(8) do
not vary significantly on the time scale of $\tau_{c}$. This assumption
is consistent with scale separation between the mean and fluctuating
quantities in the mean-field magneto hydrodynamics. The reader can
find a comprehensive discussion of the $\tau$ -approximation in the
above cited papers. Furthermore, we restrict ourselves to the high
Reynolds numbers limit and discard the microscopic diffusion terms.
The contributions of the mean magnetic field in the turbulent stresses
will be neglected because they give the nonlinear terms in the cross helicity
tensor. Also, $\tau_{c}$ is independent on $\bm{k}$ (cf, \citealt{rad-kle-rog,kle-rog:04a,2005PhR...417....1B})
and it is independent on the mean fields as well. This should be taken
into account in considering the nonlinear effects due to rotation.
Taking all the above assumptions into account, we get the system of
equations for the moments for the stationary case:

\begin{eqnarray}
\frac{\left\langle \hat{u}_{i}\left(\bm{z}\right)\hat{b}_{j}\left(\bm{z}'\right)\right\rangle }{\tau_{c}} & = & -2\left(\bm{\varOmega}\widehat{\bm{z}}\right)\varepsilon_{iln}\widehat{z_{l}}\left\langle \hat{u}_{n}(\bm{z})\hat{b}_{j}(\bm{z'})\right\rangle +i\, z'_{l}\bar{B}_{l}\hat{u}_{i}\left(\bm{z}\right)\hat{u}_{j}\left(\bm{z')}\right\rangle \label{eq:kappa1-1}\\
 & + & i\, z'_{l}\int\left[\left\langle \hat{u}_{i}\left(\bm{z}\right)\hat{b}_{l}\left(\bm{z'-q)}\right\rangle \widehat{\bar{V}}_{j}(\bm{q})-\left\langle \hat{u}_{i}\left(\bm{z}\right)\hat{b}_{j}\left(\bm{z'-q)}\right\rangle \widehat{\bar{V}}_{l}(\bm{q})\right]d\bm{q}\right.\right.\nonumber \\
 & - & i\,\pi_{if}(\bm{z)}z_{l}\int\left[\left\langle \hat{u}_{l}\left(\bm{z-q)}\hat{b}_{j}\left(\bm{z}'\right)\right\rangle \widehat{\bar{V}}_{f}(\bm{q})+\left\langle \hat{u}_{f}\left(\bm{z-q)}\hat{b}_{j}\left(\bm{z}'\right)\right\rangle \widehat{\bar{V}}_{l}(\bm{q})\right]d\bm{q}\right.\right.\nonumber \\
 & + & \frac{i}{\mu}\bar{B}_{l}z_{l}\langle\hat{b}_{i}\left(\bm{\left.z\right)}\hat{b}_{j}\left(\bm{z}'\right)\rangle,\right.\nonumber \\
\frac{\left\langle \hat{m}_{i}\left(\bm{z}\right)\hat{m}_{j}\left(\bm{z}'\right)\right\rangle }{\tau_{c}} & = & -2\left(\bm{\varOmega}\widehat{\bm{z}}\right)\varepsilon_{iln}\widehat{z_{l}}\left\langle \hat{u}_{n}(\bm{z})\hat{u}_{j}(\bm{z'})\right\rangle -2\left(\bm{\varOmega}\widehat{\bm{z}}'\right)\varepsilon_{jln}\widehat{z_{l}'}\left\langle \hat{u}_{i}(\bm{z})\hat{u}_{n}(\bm{z'})\right\rangle \label{secm1-1}\\
 & - & i\,\pi_{if}(\bm{z)}z_{l}\int\left[\left\langle \hat{u}_{l}\left(\bm{z-q)}\hat{u}_{j}\left(\bm{z}'\right)\right\rangle \widehat{\bar{V}}_{f}(\bm{q})+\left\langle \hat{u}_{f}\left(\bm{z-q)}\hat{u}_{j}\left(\bm{z}'\right)\right\rangle \widehat{\bar{V}}_{l}(\bm{q})\right]d\bm{q}\right.\right.\nonumber \\
 & - & i\,\pi_{jf}(\bm{z')}z'_{l}\int\left[\left\langle \hat{u}_{i}\left(\bm{z}\right)\hat{u}_{l}\left(\bm{z-q)}\right\rangle \widehat{\bar{V}}_{f}(\bm{q})+\left\langle \hat{u}_{i}\left(\bm{z}\right)\hat{u}_{f}\left(\bm{z-q)}\right\rangle \widehat{\bar{V}}_{l}(\bm{q})\right]d\bm{q}\right.\right.\nonumber \\
 & + & \frac{\left\langle \hat{m}_{i}\left(\bm{z}\right)\hat{m}_{j}\left(\bm{z}'\right)\right\rangle ^{\left(0\right)}}{\tau_{c}}\nonumber \\
\frac{\left\langle \hat{b}_{i}\left(\bm{z}\right)\hat{b}_{j}\left(\bm{z}'\right)\right\rangle }{\tau_{c}} & = & \frac{\left\langle \hat{b}_{i}\left(\bm{z}\right)\hat{b}_{j}\left(\bm{z}'\right)\right\rangle ^{(0)}}{\tau_{c}}\label{eq:mag1-1}\\
 & + & i\, z'_{l}\int\left[\left\langle \hat{b}_{i}\left(\bm{z}\right)\hat{b}_{l}\left(\bm{z'-q)}\right\rangle \widehat{\bar{V}}_{j}(\bm{q})-\left\langle \hat{b}_{i}\left(\bm{z}\right)\hat{b}_{j}\left(\bm{z'-q)}\right\rangle \widehat{\bar{V}}_{l}(\bm{q})\right]d\bm{q},\right.\right.\nonumber 
\end{eqnarray}
To proceed further, we have to introduce some conventions and notations
that are widely used in the literature. The double Fourier transformation
of an ensemble average of two fluctuating quantities, say $f$ and
$g$, taken at equal times and at the different positions $\bm{x},\hspace{0.25em}\bm{x}'$,
is given by 
\begin{equation}
\left\langle f\left(\bm{x}\right)g\left(\bm{x}'\right)\right\rangle =\int\int\left\langle \hat{f}\left(\bm{z}\right)\hat{g}\left(\bm{z}'\right)\right\rangle e^{i\,\left(\bm{z}\ccdot\bm{x}+\bm{z}'\ccdot\bm{x}'\right)}\mathd^{3}\bm{z}\mathd^{3}\bm{z}'.\label{cor1}
\end{equation}
 In the spirit of the general formalism of the two-scale approximation
{\citep{rob-saw}} we introduce ``fast'' and ``slow'' variables.
They are defined by the relative $\bm{r}=\bm{x}-\bm{x}'$
and the mean $\bm{R}=\frac{1}{2}\left(\bm{x}+\bm{x}'\right)$
coordinates, respectively. Then, Eq.~(\ref{cor1}) can be written
in the form 
\begin{eqnarray}
\left\langle f\left(\bm{x}\right)g\left(\bm{x}'\right)\right\rangle  & = & \int\int\left\langle \hat{f}\left(\bm{k}+\frac{1}{2}\bm{K}\right)\hat{g}\left(-\bm{k}+\frac{1}{2}\bm{K}\right)\right\rangle \mathe^{i\,\left(\bm{K}\ccdot\bm{R}+\bm{k}\ccdot\bm{r}\right)}\mathd^{3}\bm{K}\mathd^{3}\bm{k},\label{cor1b}
\end{eqnarray}
where we have introduced the wave vectors $\bm{k}=\frac{1}{2}\left(\bm{z}-\bm{z}'\right)$
and $\bm{K}=\bm{z}+\bm{z}'$. Then, following to \cite{sub-bra:04}, we
define the correlation function of $\widehat{\bm{f}}$ and $\widehat{\bm{g}}$
obtained from (\ref{cor1b}) by integration with respect to $\bm{K}$,
\begin{equation}
\Phi\left(\hat{f},\hat{g},\bm{k},\bm{R}\right)=\int\left\langle \hat{f}\left(\bm{k}+\frac{1}{2}\bm{K}\right)\hat{g}\left(-\bm{k}+\frac{1}{2}\bm{K}\right)\right\rangle \mathe^{i\,\left(\bm{K}\ccdot\bm{R}\right)}\mathd^{3}\bm{K}.\label{eq:cor1a}
\end{equation}
 For further convenience we define the second order correlations of
momentum density, magnetic fluctuations and the cross-correlations
of momentum and magnetic fluctuations via 
\begin{eqnarray}
\hat{v}_{ij}\left(\bm{k},\bm{R}\right) & = & \Phi(\hat{u}_{i},\hat{u}_{j},\bm{k},\bm{R}),\left\langle u^{2}\right\rangle \left(\bm{R}\right)=\int\hat{v}_{ii}\left(\bm{k},\bm{R}\right)\mathd^{3}\bm{k},\\
\hat{h}_{ij}\left(\bm{k},\bm{R}\right) & = & \Phi(\hat{b}_{i},\hat{b}_{j},\bm{k},\bm{R}),\left\langle b^{2}\right\rangle \left(\bm{R}\right)=\int\hat{h}_{ii}\left(\bm{k},\bm{R}\right)\mathd^{3}\bm{k},\label{cor2}\\
\hat{\varkappa}_{ij}\left(\bm{k},\bm{R}\right) & = & \Phi(\hat{u}_{i},\hat{b}_{j},\bm{k},\bm{R}),\mathcal{E}_{i}\left(\bm{R}\right)=\varepsilon_{ijk}\int\hat{\varkappa}_{jk}\left(\bm{k},\bm{R}\right)\mathd^{3}\bm{k}.
\end{eqnarray}
We now return to equations (\ref{eq:kappa1-1}), (\ref{secm1-1})
and (\ref{eq:mag1-1}). As the first step, we solve these equations
about $\boldsymbol{\varOmega}$ (non-linear effects of the Coriolis force)
and make the Taylor expansion with respect to the ``slow'' variables
and take the Fourier transformation, (\ref{eq:cor1a}), about them.
The details of this procedure can be found in  \citep{sub-bra:04}. In result we get
the following equations for the second moments

\begin{eqnarray}
\frac{\hat{\varkappa}_{ij}}{\tau_{c}} & = & -\iota\mathcal{D}_{if}^{\left(0\right)}\left(\bm{B}\ccdot\bm{k}\right)\left(v_{fj}-\frac{m_{fj}}{\mu}\right)+\mathcal{D}_{if}^{\left(0\right)}\bar{V}_{j,l}\hat{\varkappa}_{fl}-\mathcal{D}_{if}^{\left(0\right)}\bar{V}_{f,l}\hat{\varkappa}_{lj}+\label{eq:crht}\\
 & + & 2\mathcal{D}_{ip}^{\left(0\right)}\hat{k}_{p}\hat{k}_{f}\hat{\varkappa}_{lj}\bar{V}_{f,l}+\mathcal{D}_{if}^{\left(0\right)}k_{l}\bar{V}_{f,l}\frac{\partial\hat{\varkappa}_{fj}}{\partial k_{f}},\nonumber \\
\mathcal{D}_{if}^{\left(0\right)} & = & \frac{\delta_{if}+\psi_{\varOmega}\hat{k}_{p}\varepsilon_{ifp}+\psi_{\varOmega}^{2}\hat{k}_{i}\hat{k}_{f}}{1+\psi_{\varOmega}^{2}},\,\psi_{\varOmega}=2\left(\bm{\boldsymbol{\varOmega}\ccdot}\hat{\bm{k}}\right)\tau_{c}\nonumber 
\end{eqnarray}

\begin{eqnarray}
\frac{\hat{v}_{ij}}{\tau_{c}} & = & T_{ijnm}^{(0)}\left(\frac{\hat{v}_{nm}^{(0)}}{\tau_{c}}+2\hat{k}_{f}\bar{V}_{f,l}\left(\hat{k}_{n}\hat{v}_{lm}+\hat{k}_{m}\hat{v}_{nl}\right)-\bar{V}_{n,l}\hat{v}_{lm}-\bar{V}_{m,l}\hat{v}_{nl}+k_{l}\bar{V}_{f,l}\frac{\partial\hat{v}_{nm}}{\partial k_{f}}\right),\label{eq:stressS}\\
T_{ijnm}^{\left(0\right)} & = & \delta_{in}\delta_{jm}+\frac{\psi_{\varOmega}\hat{k}_{p}}{M}\left(\varepsilon_{inp}\delta_{jm}+\varepsilon_{jmp}\delta_{in}\right)\nonumber \\
 & - & \frac{\psi_{\varOmega}^{2}}{M}\left(\delta_{ij}\pi_{nm}-\delta_{nm}\hat{k}_{i}\hat{k}_{j}+\delta_{im}\hat{k}_{n}\hat{k}_{j}+\delta_{nj}\hat{k}_{i}\hat{k}_{m}-2\delta_{n\left[i\right.}\delta_{\left.j\right]m}\right),\nonumber \\
M & = & 1+4\psi_{\varOmega}^{2}\nonumber 
\end{eqnarray}
\begin{eqnarray}
\hat{h}_{ij} & = & \hat{h}_{ij}^{\left(0\right)}+\tau_{c}\hat{h}_{il}\bar{V}_{j,l}+\tau_{c}\hat{h}_{lj}\bar{V}_{i,l}+\tau_{c}k_{l}\bar{V}_{f,l}\frac{\partial\hat{h}_{ij}}{\partial k_{f}}\label{eq:secm2b}
\end{eqnarray}
These equations were solved with respect to the shear tensor, $\overline{V}_{i,j}=\nabla_{j}\overline{V}_{i}$,
by means of perturbation procedure. One remains to define the spectra
of the background turbulence. We will adopt the isotropic form of
the spectra \citep{rob-saw}. Additionally, the background magnetic
fluctuations are helical while there is no prescribed kinetic helicity
in the background turbulence: 
\begin{eqnarray}
\hat{v}_{ij}^{(0)} & = & \left\{ \pi_{ij}\left(\bm{k}\right)\frac{E\left(k,\bm{R}\right)}{8\pi k^{2}}-i\,\varepsilon_{ijp}k_{p}\frac{\mathcal{H}\left(k,\bm{R}\right)}{8\pi k^{4}}\right\} ,\label{eq:spectr1}\\
\hat{h}_{ij}^{(0)} & = & \left\{ \pi_{ij}\left(\bm{k}\right)\frac{\mathcal{B}\left(k,\bm{R}\right)}{8\pi k^{2}}-i\,\varepsilon_{ijp}k_{p}\frac{\mathcal{N}\left(k,\bm{R}\right)}{8\pi k^{4}}\right\} ,\label{eq:spectr2}
\end{eqnarray}
where, the spectral functions $E(k,\bm{R}),\mathcal{B}(k,\bm{R}),\mathcal{N}(k,R)$
define, respectively, the intensity of the velocity fluctuations,
the intensity of the magnetic fluctuations and amount of current helicity
in the background turbulence. They are defined via 
\begin{eqnarray}
\left\langle u^{(0)2}\right\rangle  & = & \int\frac{E\left(k,\bm{R}\right)}{4\pi k^{2}}\mathd^{3}\bm{k},\hspace{0.25em}\hspace{0.25em}\left\langle b^{(0)2}\right\rangle =\int\frac{\mathcal{B}\left(k,\bm{R}\right)}{4\pi k^{2}}\mathd^{3}\bm{k},\label{eq:spectr3}\\
h_{\mathcal{K}}^{\left(0\right)} & = & \hspace{0.25em}\hspace{0.25em}\int\frac{\mathcal{H}\left(k,\bm{R}\right)}{4\pi k^{2}}\mathd^{3}\bm{k},\hspace{0.25em}\hspace{0.25em}h_{C}^{\left(0\right)}=\frac{1}{\mu\rho}\int\frac{\mathcal{N}\left(k,\bm{R}\right)}{4\pi k^{2}}\mathd^{3}\bm{k},\nonumber 
\end{eqnarray}
 where $h_{\mathcal{K}}^{\left(0\right)}=\left\langle \bm{u^{\left(0\right)}\ccdot\nabla\times u^{\left(0\right)}}\right\rangle $
and $h_{\mathcal{C}}^{\left(0\right)}=\frac{\left\langle \bm{b^{\left(0\right)}\ccdot\nabla\times b^{\left(0\right)}}\right\rangle }{\mu\rho}$.
In final results we use the relation between intensities of magnetic
and kinetic fluctuations which is defined via $\mathcal{B}\left(k,\bm{R}\right)=\varepsilon\mu\bar{\rho}E\left(k,\bm{R}\right)$.
The state with $\varepsilon=1$ means equipartition between energies
of magnetic an kinetic fluctuations in the background turbulence.

\section*{Appendix B}

\begin{align*}
f_{1}^{(\gamma)} & =\frac{1}{(24\varOmega^{*})^{2}}\left(\left(1300\varOmega^{*2}+391\right)\frac{\arctan\left(2\varOmega^{*}\right)}{2\varOmega^{*}}-1456\left(\varOmega^{*2}+1\right)\frac{\arctan\left(\varOmega^{*}\right)}{\varOmega^{*}}-3(32\varOmega^{*2}-355)\right)\\
f_{2}^{(\gamma)} & =\frac{3}{4\varOmega^{*2}}\left((\varOmega^{*2}+1)\frac{\arctan\left(\varOmega^{*}\right)}{\varOmega^{*}}-1\right)\\
f_{3}^{(\gamma)} & =-\frac{1}{36540\varOmega^{*}{}^{4}}\left(5\left(\left(10672\varOmega^{*2}+3872\right)\varOmega^{*2}+337\right)\frac{\arctan\left(2\varOmega^{*}\right)}{2\varOmega^{*}}\right.\\
 & -\left.320\left(\left(515\varOmega^{*2}-128\right)\varOmega^{*2}-895\right)\frac{\arctan\left(\varOmega^{*}\right)}{\varOmega^{*}}+3\left(\left(2304\varOmega^{*2}+3380\right)\varOmega^{*2}-48295\right)\right)\\
f_{4}^{(\gamma)} & =-\frac{1}{24\varOmega^{*4}}\left(3\left(\left(11\varOmega^{*2}+8\right)\varOmega^{*2}-7\right))\frac{\arctan\left(\varOmega^{*}\right)}{\varOmega^{*}}-\left(31\varOmega^{*2}-21\right)\right)\\
f_{5}^{(\gamma)} & =-\frac{1}{55296\varOmega^{*4}}\left(\left(7472\varOmega^{*4}+5016\varOmega^{*2}-1685\right)\frac{\arctan\left(2\varOmega^{*}\right)}{2\varOmega^{*}}\right.\\
 & -32(79\varOmega^{*4}+1410\varOmega^{*2}+4475)\frac{\arctan\left(\varOmega^{*}\right)}{\varOmega^{*}}\\
 & +\left.3\frac{\left(9872\varOmega^{*6}+183632\varOmega^{*4}+238183\varOmega^{*2}+48295\right)}{\left(\varOmega^{*2}+1\right)\left(4\varOmega^{*2}+1\right)}\right)\\
f_{6}^{(\gamma)} & =\frac{1}{32\varOmega^{*4}}\left(\left(\left(9\varOmega^{*2}+30\right)\varOmega^{*2}-35\right)\frac{\arctan\left(\varOmega^{*}\right)}{\varOmega^{*}}-\frac{\left(\left(101\varOmega^{*2}+20\right)\varOmega^{*2}-105\right)}{3\left(\varOmega^{*2}+1\right)}\right)\\
f_{7}^{(\gamma)} & =-\frac{1}{55296\varOmega^{*4}}\left(\left(20528\varOmega^{*4}+16536\varOmega^{*2}-1685\right)\frac{\arctan\left(2\varOmega^{*}\right)}{2\varOmega^{*}}\right.\\
 & -32\left(577\varOmega^{*4}+3660\varOmega^{*2}+4475\right)\frac{\arctan\left(\varOmega^{*}\right)}{\varOmega^{*}}\\
 & +\left.3\frac{\left(80528\varOmega^{*6}+282512\varOmega^{*4}+258343\varOmega^{*2}+48295\right)}{(\varOmega^{*2}+1)(4\varOmega^{*2}+1)}\right)\\
f_{8}^{(\gamma)} & =-\frac{(\varOmega^{*4}+7)}{32\varOmega^{*4}}\left((\varOmega^{*2}+5)\frac{\arctan\left(\varOmega^{*}\right)}{\varOmega^{*}}-\frac{\left(13\varOmega^{*2}+15\right)}{3(\varOmega^{*2}+1)}\right)
\end{align*}

\end{document}